\documentclass[journal=jpccck,manuscript=article]{achemso}

\usepackage[version=3]{mhchem} % Formula subscripts using \ce{}
\usepackage{siunitx}
\usepackage{chemmacros}
\usepackage{enumerate}

\usepackage{hyperref}
\usepackage{caption}
\usepackage{subcaption}
\newcommand{\fig}[1]{Figure\  \ref{#1}}
\newcommand{\figtwo}[2]{Figure\ \ref{#1} (\subref{#2})}

\DeclareSIUnit{\calorie}{cal}
\DeclareSIUnit{\atmosphere}{atm}

\author{Parham Rezaee}
\altaffiliation{\textup{These Authors have the same contribution.}}
\affiliation{Chemistry Department, Sharif University of Technology, Tehran 11155-9516, Iran}

\author{Mahmood Akbari}
\altaffiliation{\textup{These Authors have the same contribution.}}
\affiliation{UNESCO-UNISA Africa Chair in Nanoscience \& Nanotechnology (U2ACN2), College of Graduate Studies, University of South Africa (UNISA), Pretoria, South Africa}
\alsoaffiliation{Nanoscience African Network (NANOAFNET), Material Research Division, iThemba LABS-National Research Foundation, Somerset West 7129, South Africa}

\author{Razieh Morad}
\altaffiliation{\textup{These Authors have the same contribution.}}
\affiliation{UNESCO-UNISA Africa Chair in Nanoscience \& Nanotechnology (U2ACN2), College of Graduate Studies, University of South Africa (UNISA), Pretoria, South Africa}
\alsoaffiliation{Nanoscience African Network (NANOAFNET), Material Research Division, iThemba LABS-National Research Foundation, Somerset West 7129, South Africa}

\author{Amin Koochaki}
\altaffiliation{\textup{These Authors have the same contribution.}}
\affiliation{Chemistry Department, Sharif University of Technology, Tehran 11155-9516, Iran}

\author{Malik Maaza}
\email{maaza@tlabs.ac.za}
\affiliation{UNESCO-UNISA Africa Chair in Nanoscience \& Nanotechnology (U2ACN2), College of Graduate Studies, University of South Africa (UNISA), Pretoria, South Africa}
\alsoaffiliation
{Nanoscience African Network (NANOAFNET), Material Research Division, iThemba LABS-National Research Foundation, Somerset West 7129, South Africa}

\author{Zahra Jamshidi}
\email{njamshidi@sharif.edu}
\affiliation
{Chemistry Department, Sharif University of Technology, Tehran 11155-9516, Iran}

\title {First Principle Simulation of Coated Hydroxychloroquine on Ag, Au and Pt Nanoparticle as a Potential Candidate for Treatment of SARS-CoV-2 (COVID-19)}


\SectionNumbersOn
\let\oldmaketitle\maketitle
\let\maketitle\relax

\begin{document}

%\begin{tocentry}
%\end{tocentry}

\oldmaketitle
\begin{abstract}

The {\it{in vitro}} antiviral activity of Hydroxychloroquine (HCQ) and chloroquine (CQ) against SARS-CoV-2 from the first month of pandemic proposed these drugs as the appropriate therapeutic candidate, although their side effect directed the clinical test toward optimizing the safe utilization strategies. The noble metal nanoparticles (NP) as promising materials with antiviral and antibacterial properties can deliver the drug to the target agent and decrease the side effect. In this work, we have applied quantum mechanical and classical atomistic molecular dynamics computational approaches to demonstrate the adsorption properties of HCQ on Ag, Au, AgAu, and Pt nanoparticles. The adsorption energies(less than $\SI{-30}{\kilo\calorie\per\mol}$)  were established for HCQ, and the (non)perturbative effects of this drug on the plasmonic absorption spectra of AgNP and AuNP have characterized with time-dependent density functional theory. The effect of size and compositions of nanoparticle on the coating with HCQ and CQ have obtained and proposed the appropriate candidate for drug delivery. This kind of modeling could help the experimental groups to find the efficient and safe therapies. 
\\
\begin{figure*}
\center
  \includegraphics[width=0.8\linewidth]{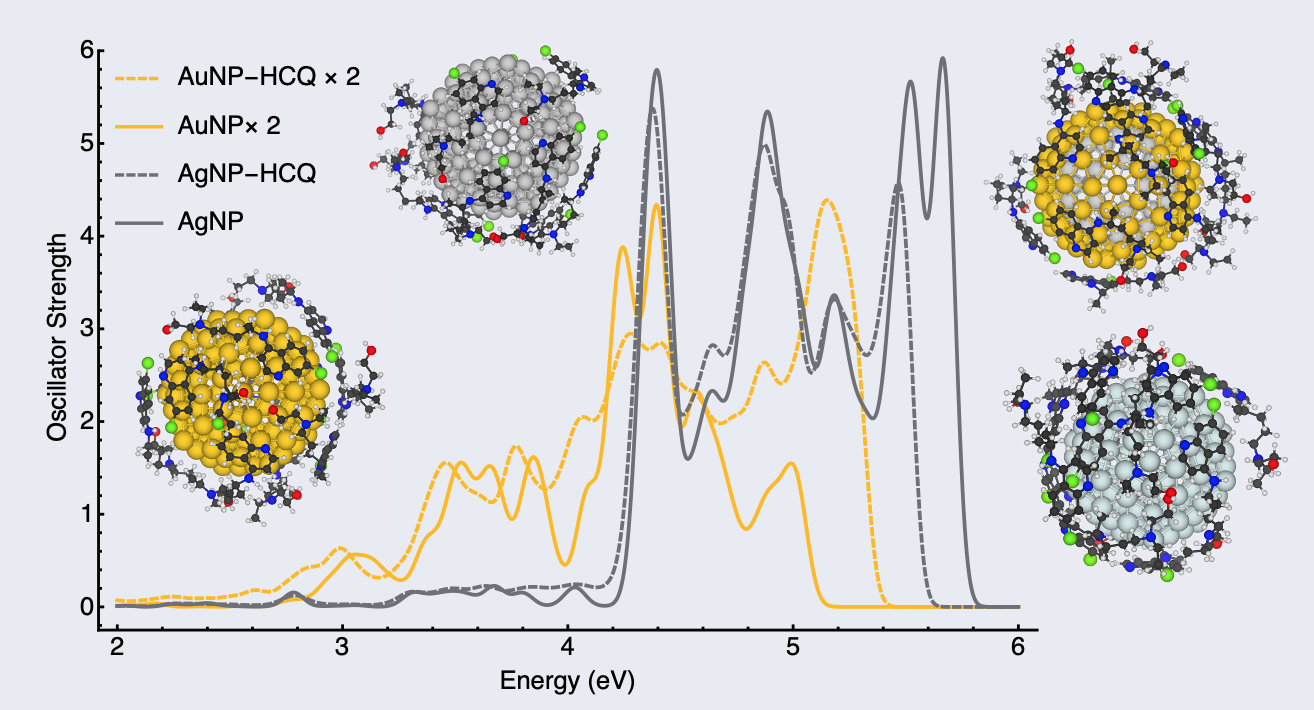}
  \label{fig_GA}
\end{figure*}

\end{abstract}



\section{Introduction}

The pandemic coronavirus disease-19 (COVID-19) has severely compromised healthcare system, and more than 300 clinical research group over the world focus on investigating the potential therapeutic options for the prevention and treatment of COVID-19\cite{noauthor_nih_2020}. Under the global health crisis, one of the most rapid and reliable treatment is drug repurposing, means examination of existing FDA approved drugs for new therapeutic purposes\cite{shah_guide_2020}. As development of new drug into clinical candidate takes approximately 10 years, and vaccine development process will take 12 to 18 months. However, the genetic sequence similarity of syndrome coronavirus, SARC-CoV, to SARC-CoV-2 (or COVID-19) with nearly same structure in viral targets that might be inhibited by the same compounds makes the possibility of usage the FDA approved antiviral drug for treatment. Chloroquine (CQ) and hydroxychloroquine (HCQ) have been used for many years as pharmacotherapies for malaria. It is believed that these drugs inhibit SARS-CoV infection and have been labeled as a potential therapeutic option against COVID-19\cite{liu_hydroxychloroquine_2020}.

The pre-clinical {\it{in vitro}} studies have done and showed the prophylactic and antiviral effects of CQ and HCQ against SARS-CoV-2\cite{wang_remdesivir_2020, yao_vitro_2020, gautret_hydroxychloroquine_2020}. On the other hand, the recent clinical outcomes opinions advocated effect in preventing and treating COVID-19, but to make a clear conclusion and optimizing safe therapeutic strategies, more examinations are desperately needed\cite{giudicessi_urgent_2020, holshue_first_2020}. It was established that these drugs by increasing the pH affects the viral replication\cite{shah_systematic_2020}. However, the mechanism of action of HCQ/CQ against COVID-19 has not yet fully elucidated, and it was known HCQ shares the same mechanism of action as CQ, and it is less toxic\cite{liu_hydroxychloroquine_2020}. However, a large scale and prolonged usage seems potentially dangerous and increased the risk of drug-induced torsades de pointes and may lead to cardiac death \cite{lecuit_chloroquine_2020}. Therefore, different treatment regimens try to focus on efficient {\it{in vivo}} usage of these drugs\cite{colson_chloroquine_2020, pastick_review:_2020}. 

In this regards, nanomedicine as growing field of nanotechnology has a wide-spread application in pharmaceuticals and diagnostic devices\cite{rai_strategic_2015, jeyaraj_comprehensive_2019}. Metal nanoparticles are well-known as a promising material that able to transport drugs to specific targets in the body, and could be engineered to develop new drug delivery systems\cite{ravindra_development_2012}. Noble metals have been used as therapeutic agents from the ancient time in medicine. Noble metal nanoparticles particularly, silver, gold and platinum nanoparticles are revealing the stability in biological environment and survive in an intracellular\cite{burdusel_biomedical_2018, pedone_platinum_2017, pena-gonzalez_dendronized_2016}. The stable nanoparticles with small size reveal a benefit that straightforwardly interact with biomolecules both at surface and inside cells, and play major role in biomedical applications such as drug vehicle in diagnosis and treatment of diseases. Silver has a long history of its usage as antibacterial properties\cite{geraldo_green_2016, chowdhury_chocolate_2016, tavaf_evaluation_2017, henke_antibacterial_2016}, and recent studies utilized the antiviral and immunomodulatory properties of silver nanoparticles (AgNPs)\cite{galdiero_silver_2011, villeret_silver_2018}. Recently, R. P. Garofalo\cite{morris_antiviral_2019} and coworker, demonstrated {\it{in vivo}} antiviral activity of AgNPs during respiratory syncytial virus (RSV) infection. 

Under extended lockdown across the world, with no possibility of accessing to experimental laboratories full-time, the hard work is just beginning for computational chemists and biophysicists to model different approaches and propose the efficient therapies to the experimentalist. In this work, the first principle density functional calculations were carried out to find out the affinity of HCQ/CQ molecules towards silver and gold nanoparticles, and confirmed their weak interaction by theoretical UV-Vis absorption spectra. The complimentary calculations by changing the size and composition of metal nanoparticle and also the number of coated HCQ molecule were done based on molecular dynamics simulations. 

\section{Computational Methods}

For HCQ and CQ molecules, the geometry optimization and frequency calculation were performed with PBE generalized gradient (GGA) exchange-correlation (xc-) density functional\cite{PBE} with inclusion of the Grimme dispersion correction scheme (D3) \cite{Grimme2011, fonseca_guerra_towards_1998, handy_left-right_2001} applying Becke-Johnson damping and a triple-$\zeta$ polarized (TZP) Slater type basis set (PBE-D3/TZP). The Conductor like Screening Model (COSMO)\cite{klamt_cosmo:_1993} was considered to model the effect of water solvent. For the optimized structure the Hirshfeld point charges\cite{schrier_carbon_2012} and electrostatic potential map were obtained with and without solvent (in \fig{fig_charge} and Figure S1 in supporting information). The experimental interatomic metal-metal distance, was employed to create starting structures for further optimization with the LDA (local density approximation) xc-functional\cite{vosko_accurate_1980} and the scalar relativistic ZORA formalism\cite{pierloot_relative_2008, van_lenthe_zeroorder_1996}. The interactions of HCQ with icosahedral silver and gold clusters with 147 atoms have been investigated at the PBE-D3/TZP level of theory under the influence of relativistic effect (ZORA). To find out the effect of HCQ drug on the electronic structures and the absorption plasmonic spectra of noble metal particle, the recently developed time-dependent density functional approach, TD-DFT+TB method\cite{ruger_tight-binding_2016, Asadi_Aghbolaghi2020} which combines a full DFT ground state with tight-binding approximations, was applied. The exited states calculations were performed at the optimized geometries by utilizing the asymptotically corrected LB94 xc-functional \cite{PhysRevA.49.2421} and obtained the absorption spectra in the range of 0.0-6.0 \si{\electronvolt}. All these calculations were performed with the Amsterdam Density Functional (ADF2019.1) program\cite{te_velde_chemistry_2001}.

The adsorption of HCQ and CQ molecules on a periodic slab model of the Ag (111), Au(111) and Pt(111) surfaces were simulated using the plane-wave based Quantum ESPRESSO Package\cite{QuantumEspresso} with PBE-D3\cite{PBE, Grimme2011} functional. A plane-wave basis set with an energy cut-off of 80 Rydberg  ($\sim$ 1088 \si{\electronvolt}) has been employed and the electron–ion interactions were represented through the ultrasoft pseudopotential including the scalar relativistic effects. Two layers of Ag/Au/Pt atoms were considered with a total number of 36 atoms per layer. A large vacuum of in the z direction (perpendicular to the bilayer) was applied and due to the large size of the box, only the  $\Gamma$-point of the reciprocal lattice was considered. During the calculations, the internal slab atoms were kept fixed at the bulk positions while the atoms in the top layer and molecule were allowed to relax.

The molecular dynamics (MD) simulations for interactions of HCQ and CQ with noble metal nanoparticle (NP), \ce{Ag147}, \ce{Au147}, \ce{Au92Ag55} and \ce{Pt147} were performed in a cubic box with sides of $\num{60 x 60 x 60} \si{\angstrom}$ (Structures are available in Table S1--S6 in supporting information). The nanoparticles fixed at the center of box which surrounded with randomly filled of the water molecules and either HCQ or CQ drugs. The time step in the simulation was \SI{2.0}{\femto\second} and the length of time is \SI{20}{\nano\second}, under the isothermal-isobaric NPT (constant particle number, pressure and temperature) condition at \SI{300}{\kelvin} and \SI{1}{\atmosphere} (controlled with Nose--Hoover algorithm). The OPLS-AA\cite{jorgensen_opls_1988} and TIP3P\cite{jorgensen_comparison_1983} force fields were used to describe the interactions of drugs and water molecules, respectively. Moreover, the Lennard-Jones parameters for nanoparticles are listed in table S7. The electrostatic interactions were simulated with the Particle-Particle Particle-Mesh (PPPM)\cite{hockney_computer_1988} solver (with accuracy \num{1e-5}). The non-bonded dispersion interactions were computed with Lennard-Jones (LJ) 12-6 potential with the cut-off distance of \SI{12}{\angstrom}. 

In order to study the effect of nanoparticle size, four different sizes of AgNPs with 147-, 561-, 1415-, 2869-atoms (with radius 1.6, 2.6, 3.6, and 4.6 \si{\nano\metre}, respectively), placed at the simulation box with fixed (12) and different number of HCQ molecules (12, 32, 64, 105) that increased proportionally with the number of silver atoms on the surface. The simulation box filled with 6000 water molecules. All simulations were performed using the Gromos53a6\cite{kyrychenko_polyvinyl_2017} and SPC\cite{jorgensen_comparison_1983} force fields. The energy minimized by using the steepest descent minimization algorithm\cite{phanchai_insight_2018}. Each system was equilibrated in both NVT ensemble coupled to the V-rescale thermal bath at \SI{300}{\kelvin} over \SI{100}{\pico\second} and in the NPT ensemble coupled to the Berendsen pressure bath at \SI{1}{\atmosphere} over \SI{200}{\pico\second}. Each system was then subjected to a \SI{50}{\nano\second} molecular dynamics (MD) simulation under constant conditions of \SI{1}{\atmosphere} and \SI{300}{\kelvin} with a time step of \SI{1}{\femto\second}. The bond lengths constrained using the LINCS algorithm\cite{Hess97lincs:a} and the long-range electrostatics using the particle mesh Ewald (PME)\cite{Plimpton97particle-meshewald} were applied. The MD simulations were carried out using the LAMMPS\cite{plimpton_fast_1995-1}, and GROMACS packages\cite{abraham_gromacs:_2015,van_der_spoel_gromacs:_2005} and VMD package \cite{Humphrey1996VMD} was used for visualization.

\section{Results and Discussion}

\begin{figure*}
  \includegraphics[width=\linewidth]{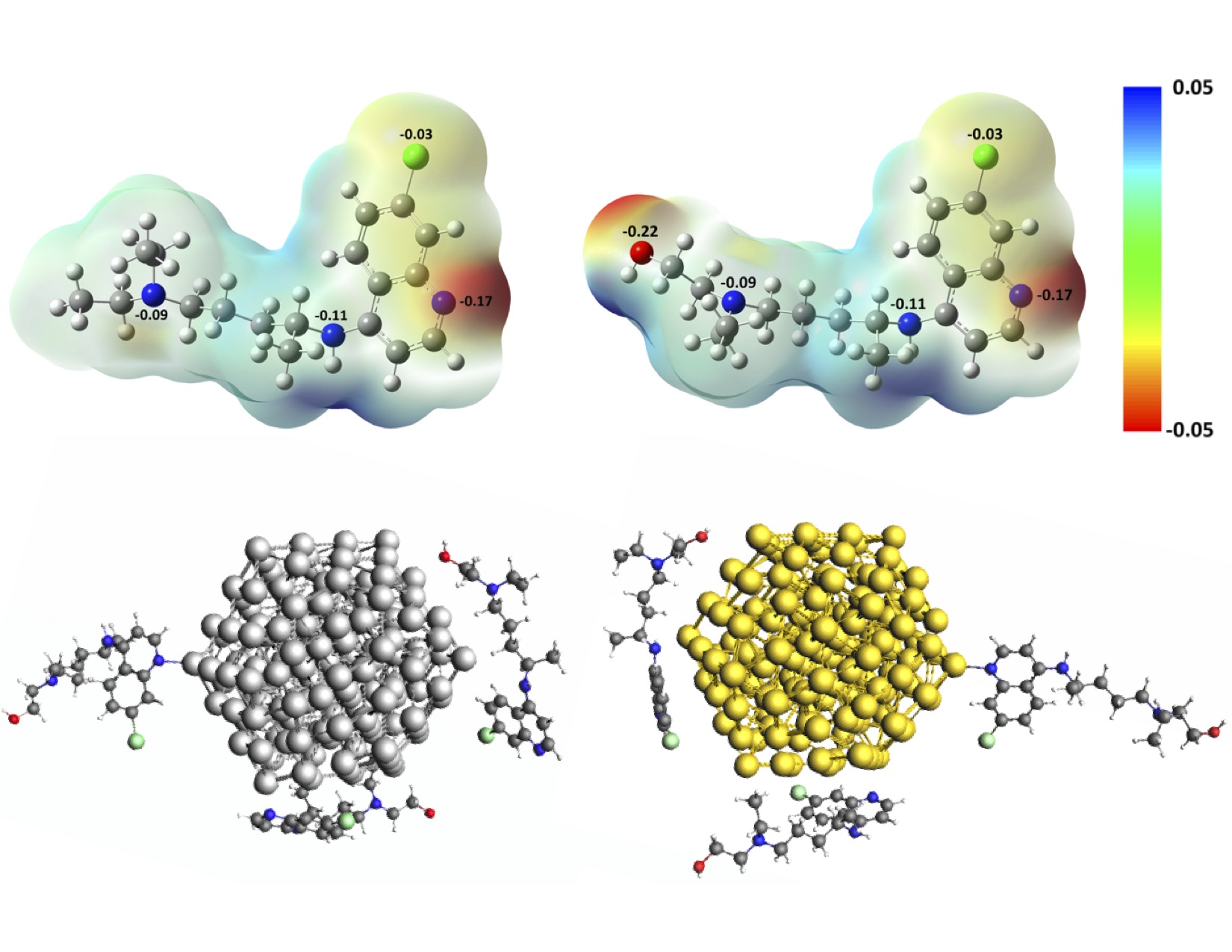}
  \caption{Charge distribution of HCQ and CQ molecules and their electrostatic potential map. Stable geometry of \ce{Ag147} and \ce{Au147} complexed with HCQ molecules (at BPE-D3/TZP level of theory).}
  \label{fig_charge}
\end{figure*}

\begin{figure*}
%=============  Ag (111)  ============
\begin{subfigure}{.43\textwidth}
  \centering
  \includegraphics[width=\linewidth]{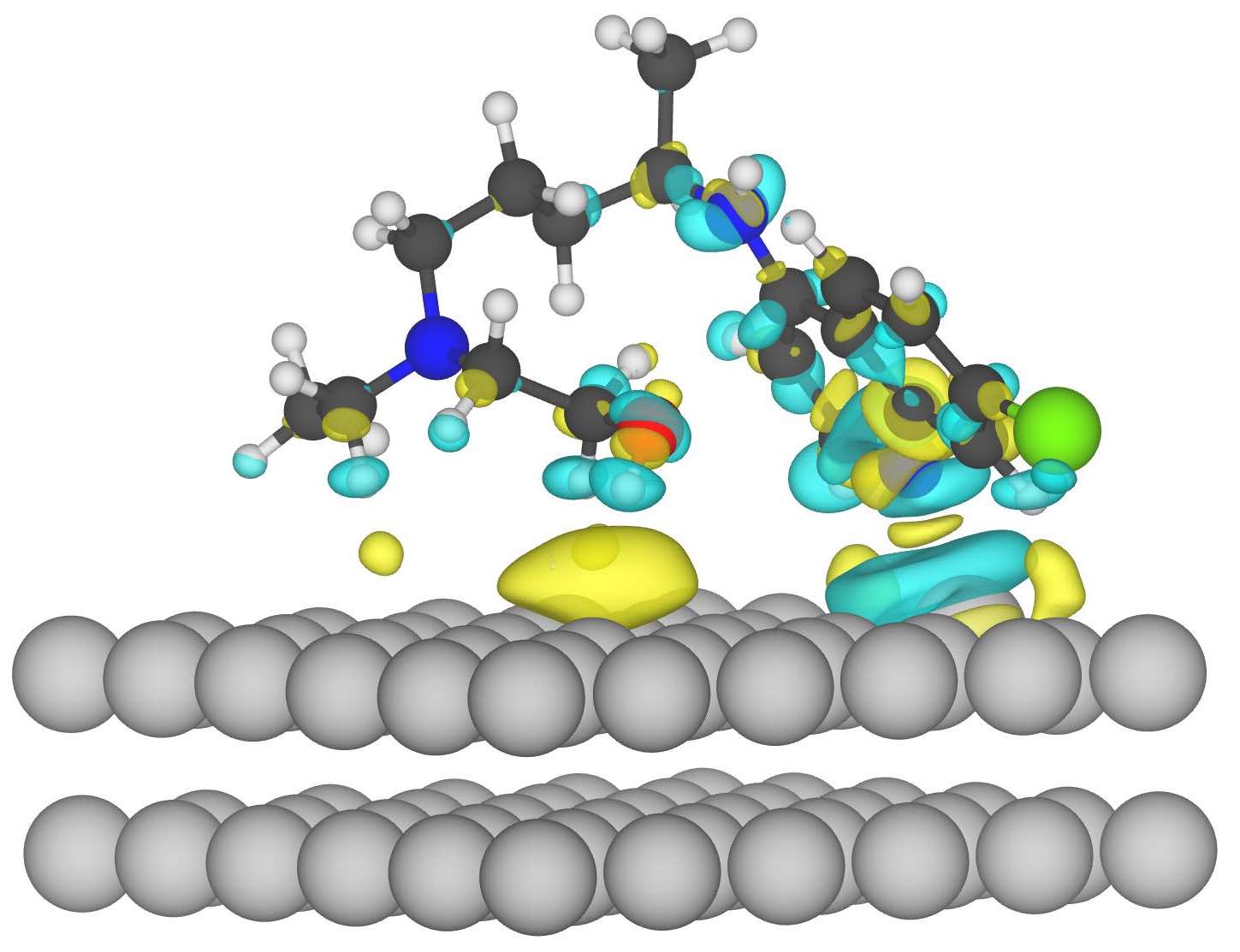}    
   \caption{}						 \label{Ag(111)HCQ}
\end{subfigure}
\begin{subfigure}{.48\textwidth}
  \centering
  \includegraphics[width=1.\linewidth]{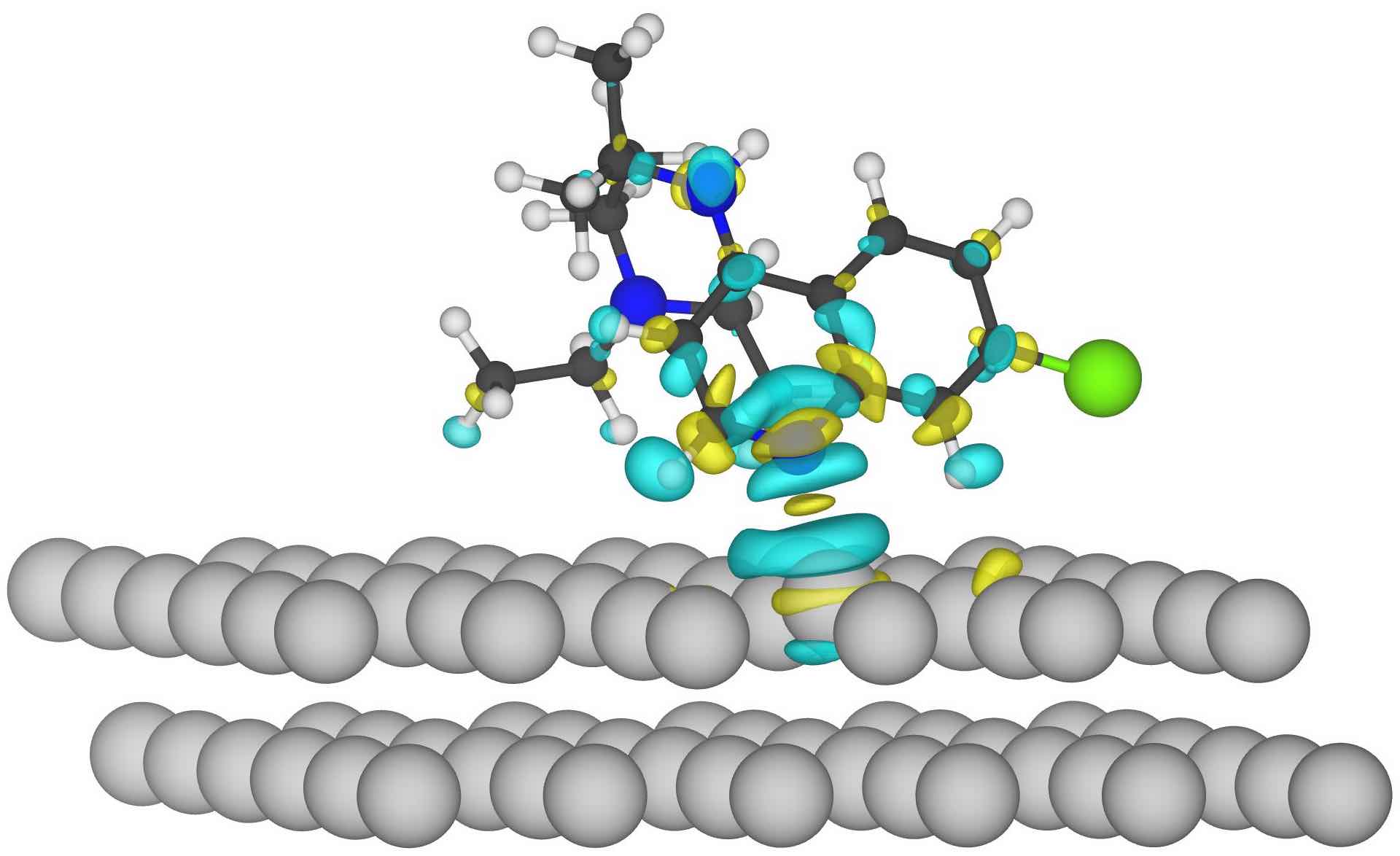}    
   \caption{}						 \label{Ag(111)CQ}
\end{subfigure}
%=============  Au (111)  ==============
\begin{subfigure}{.43\textwidth}
  \centering
  \includegraphics[width=\linewidth]{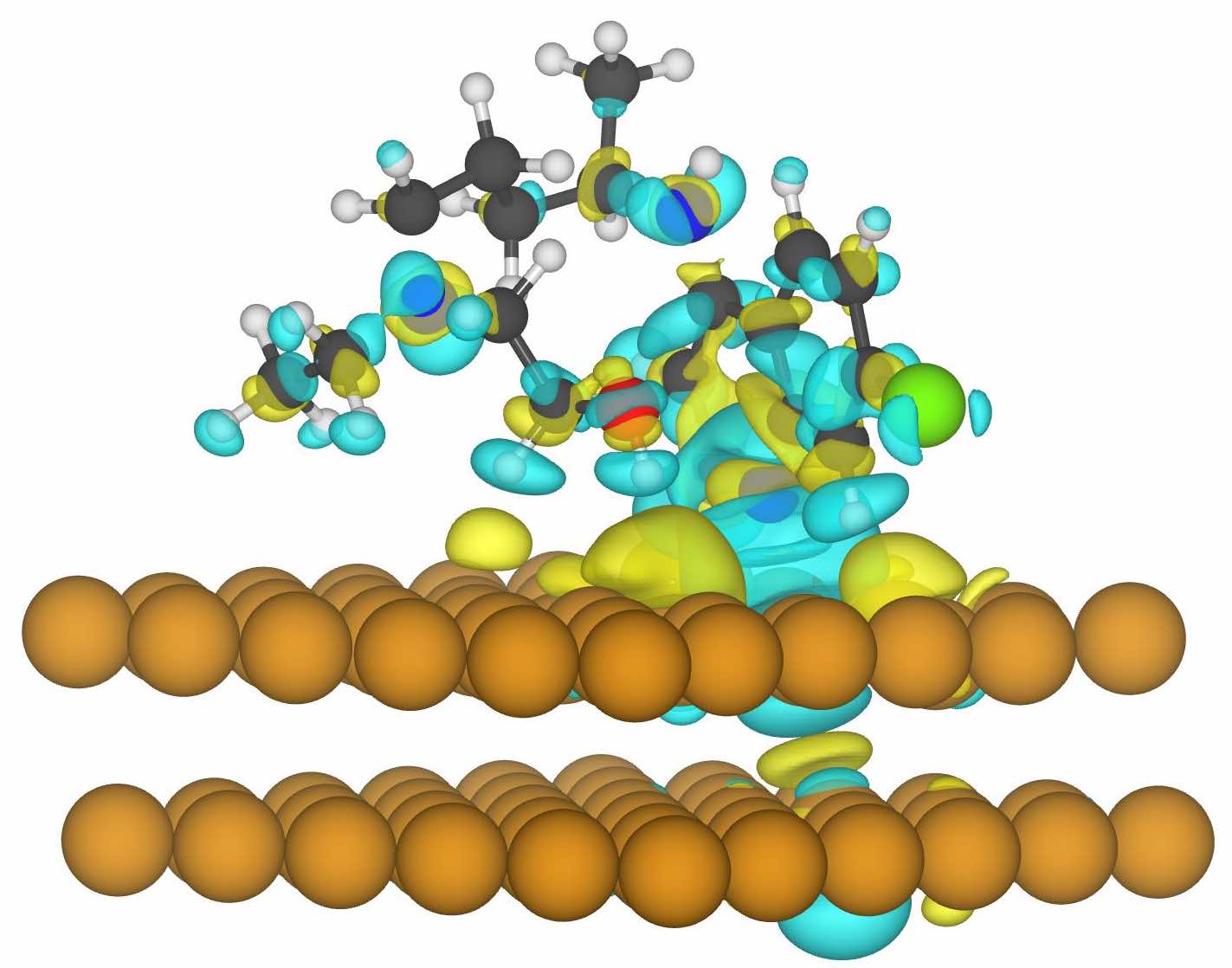}    
   \caption{}						 \label{Au(111)HCQ}
\end{subfigure}
\begin{subfigure}{.48\textwidth}
  \centering
  \includegraphics[width=\linewidth]{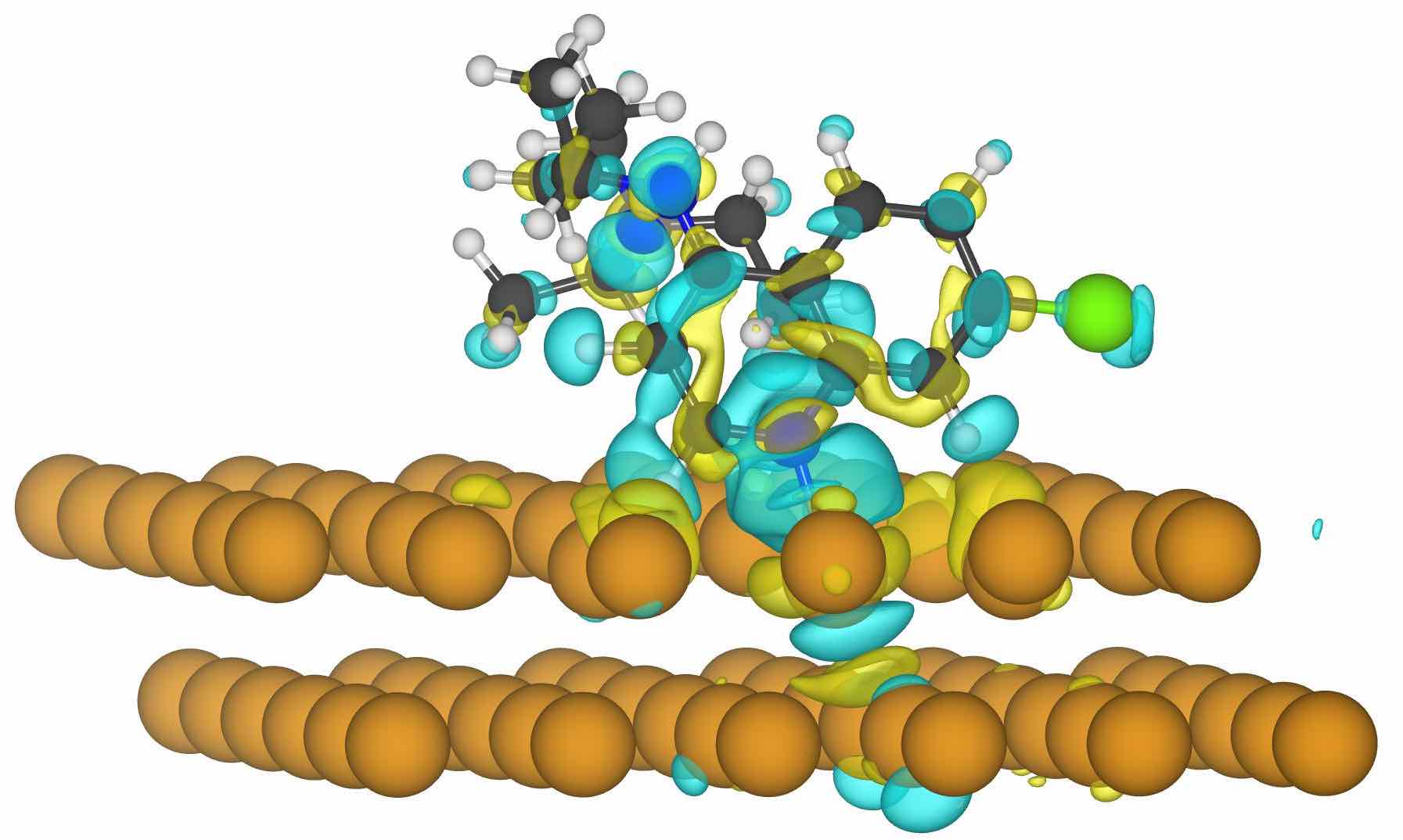}    
   \caption{}						 \label{Au(111)CQ}
\end{subfigure}
%=============  Pt (111)  ==============
\begin{subfigure}{.43\textwidth}
  \centering
  \includegraphics[width=\linewidth]{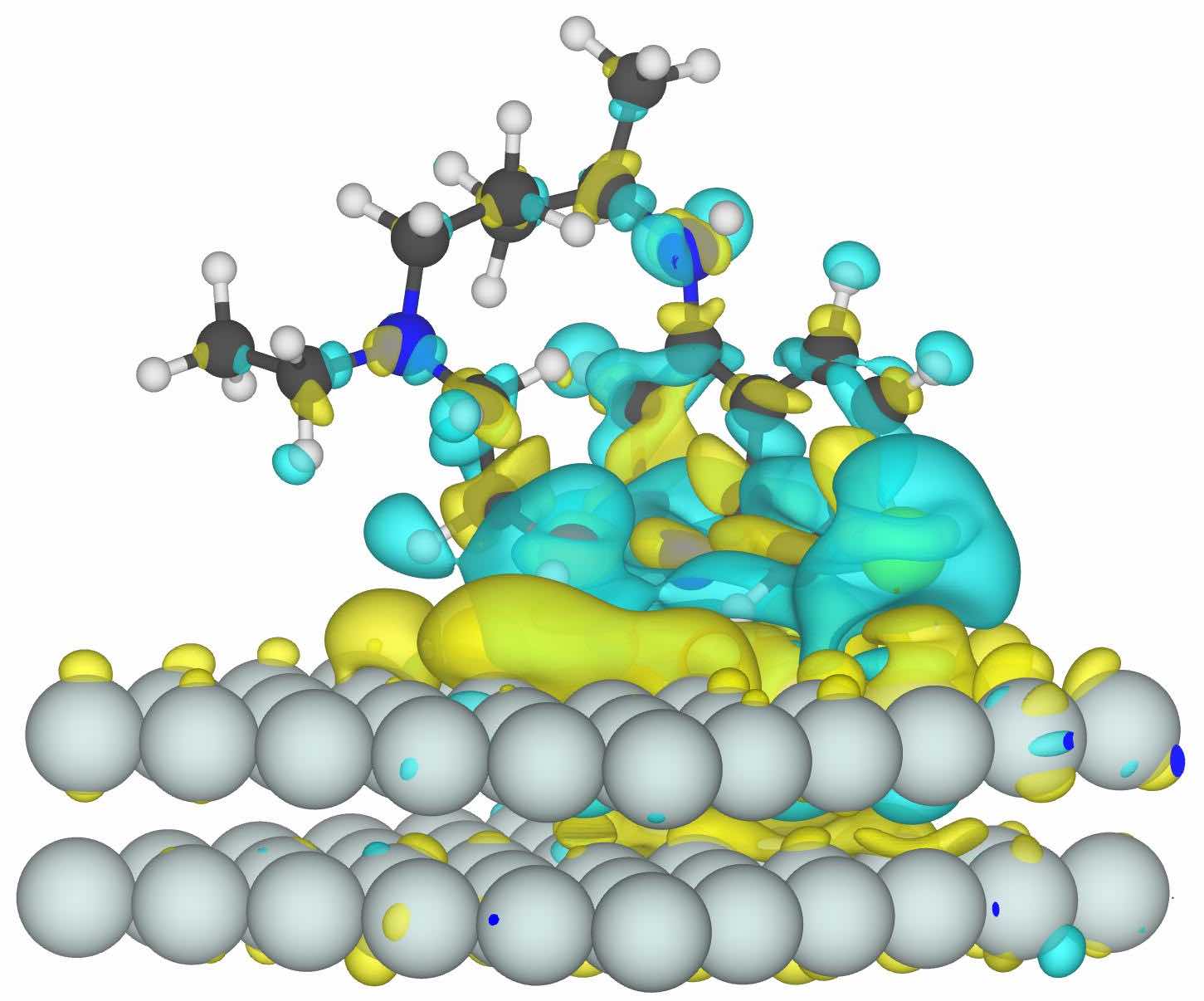}    
   \caption{}						 \label{Pt(111)HCQ}
\end{subfigure}
\begin{subfigure}{.48\textwidth}
  \centering
  \includegraphics[width=\linewidth]{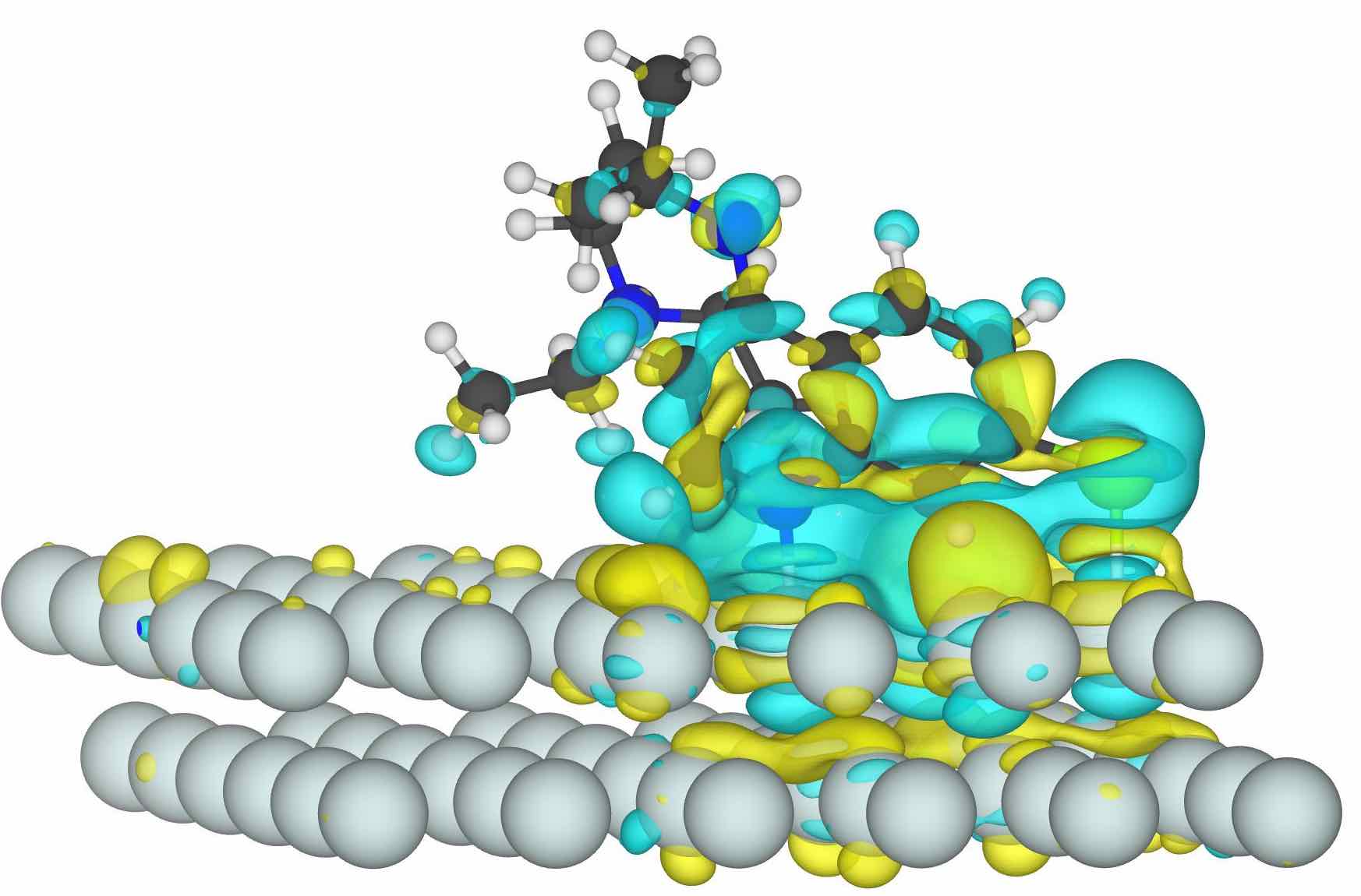}    
   \caption{}						 \label{Pt(111)CQ}
\end{subfigure}
\caption{The charge density difference of Adsorbed of HCQ /CQ molecules on a periodic slab model of the \ce{Ag (111)}, \ce{Au (111)}, and \ce{Pt (111)}. The isovalue for the charge transfer plot, is fixed at $0.001 e / a.u^3$. Yellow and blue color indicate positive and negative level corresponds to gain and loss of electron density.}
\label{fig_planewave}
\end{figure*}

\textbf{\large Interaction of AgNP and AuNP with HCQ(CQ).} The charge distribution of HCQ and CQ molecules and their electrostatic potential map in Figure \ref{fig_charge} displays the active sites of these molecules for interaction with noble metal NPs. The initial structure of complexes generated by placing the small silver cluster near the electron-rich sites (such as N-, O- and Cl- groups). These sites can donate the electron density via their lone pairs to 4d and 5s orbitals of silver atom\cite{Pakiari2007, Tehrani2012}. For the CQ and HCQ, the nitrogen of pyridine ring and (for HCQ) the oxygen of hydroxyl group have the highest affinity for interaction with noble metal clusters. 
Moreover, the optimized structure of HCQ/CQ on Ag(111), Au(111), and Pt(111) bilayer exhibited the highest affinity of drug molecules toward the platinum surface and their charge density difference confirmed the transferred of charge and accumulation on the metal surface (see \fig{fig_planewave} and also Table S8 in supporting information).

\begin{figure*}
%=============  Ag - Plasmonics  ==========================
\begin{subfigure}{.49\textwidth}
  \centering
  \includegraphics[width=\linewidth]{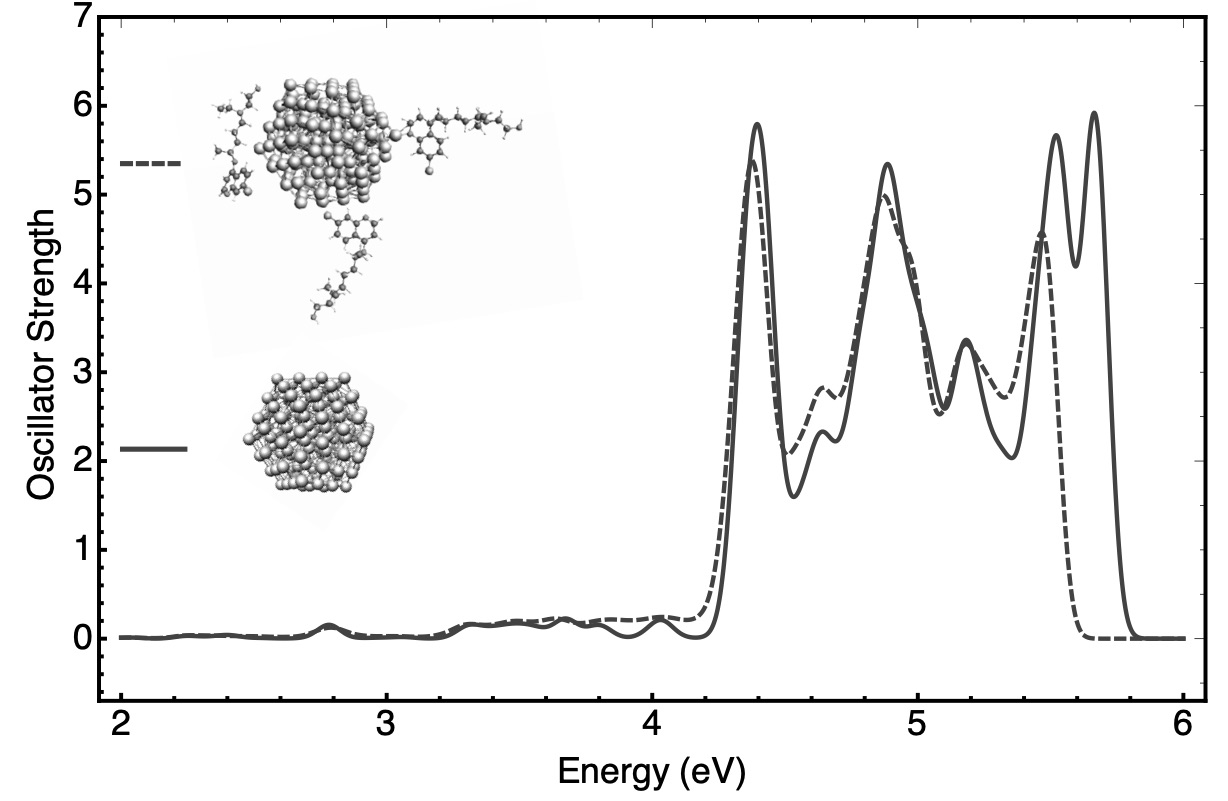}    
 \caption{}					     \label{fig_Ag_plasmonic}
\end{subfigure}
%=============   Au - Plasmonics  ==========================
\begin{subfigure}{.49\textwidth}
  \centering
  \includegraphics[width=\linewidth]{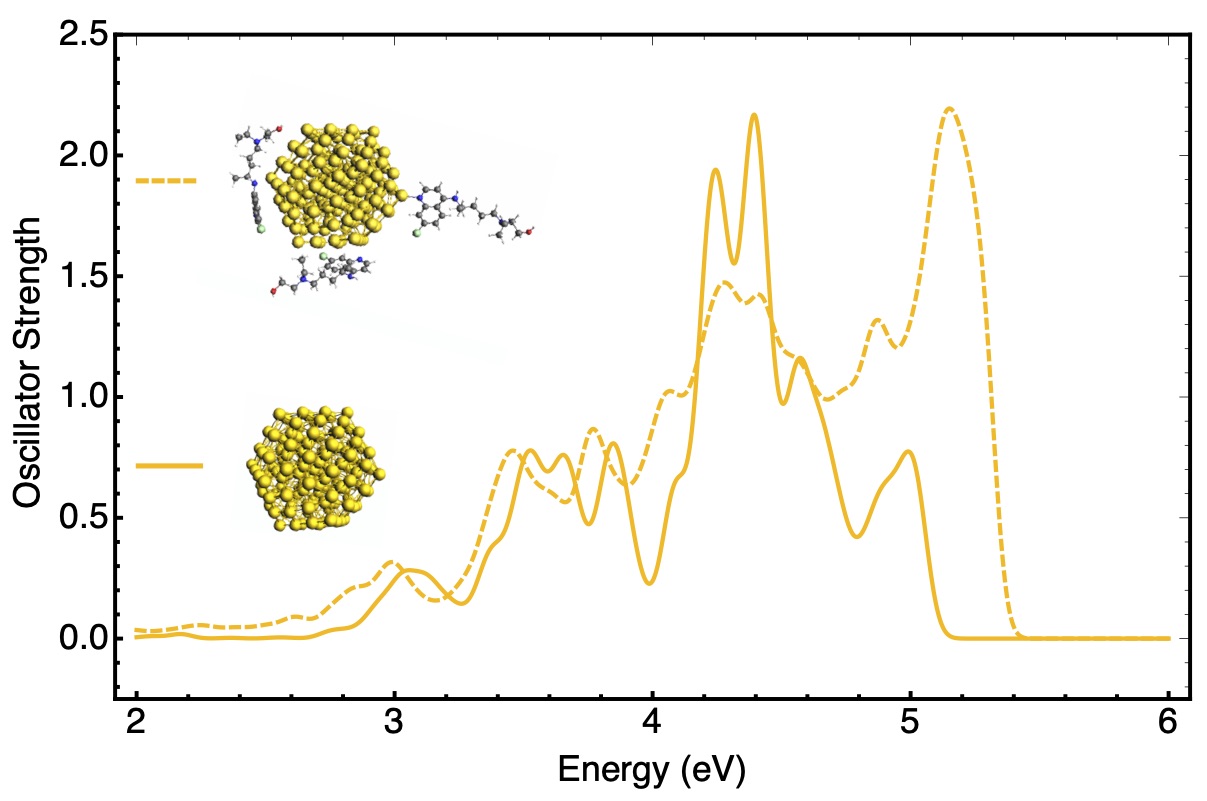}   
   \caption{}						 \label{fig_Au_plasmonic}
\end{subfigure}
\caption{Comparison the plasmonic absorption spectra of bare (solid line) and complexed (dash line) for (\subref{fig_Ag_plasmonic}) AgNP-HCQ and (\subref{fig_Au_plasmonic}) AuNP-HCQ at the TD-DFT+TB level of theory. Spectra have been broadened with a $\sigma = 0.1 $ \si{\electronvolt}  Gaussian.}\label{fig_plasmonic}
\end{figure*}

In addition,  \fig{fig_charge} shows the stable geometry of icosahedral \ce{Ag147} and \ce{Au147} nanoparticles which complexed with HCQ molecules (at PBE-D3/TZP level of theory). Herein, the non-covalent charge-transfer interactions with partially negative charge groups of the molecules plays the essential role in determining the ability of nanoparticle to catching HCQ(CQ). The binding energy of HCQ with AgNP (at PBE-D3/TZP level of theory) is about  $\Delta E_b = \SI{-21.06}{\kilo\calorie\per\mol}$, (per number of HCQ molecule,) while, for AuNP the interaction energy is about $\Delta E_b = \SI{-29.39}{\kilo\calorie\per\mol}$ that is more favorable than silver. The more electron affinity of gold ($EA_{Au} = \SI{2.31}{\electronvolt}$) with respect to silver ($EA_{Ag} = \SI{1.30}{\electronvolt}$)\cite{doi:10.1021/jp0757098} increases the interactions energy of gold atoms toward the lone-pair of HCQ, which is also confirmed with density difference map and accumulation of negative charge on the Au(111) surface. In this regard, the adsorption energy of HCQ toward Pt(111) is about 40\% more than gold surface. (see \fig{fig_planewave} and Table S8).\\
\textbf{\large Absorption Spectra of AgNP and AuNP with HCQ.} The absorption spectra of bare nanoparticle and its variation under the effect of coated compounds is another important evidence that can be compared with the experimental results, and estimate accurately the effect of adsorbent molecules on the variation of electronic structures of metal NP. Silver and gold nanoparticle are well-known with their high-intensity plasmonic absorption spectra in the range of UV-Vis that can be varied by coating with drugs. Herein, the TD-DFT+TB calculations for the optimized AgNP-HCQ and AuNP-HCQ complexes in comparison to bare nanoparticles were obtained. TD-DFT+TB as an accurate and efficient approach, obtained the ground-state orbitals with DFT and excited-state properties with tight-binding method. As can be found in \fig{fig_plasmonic}, for AgNP-HCQ complex, the plasmonic spectrum in comparison to bare AgNP did not change clearly and just the intensity of peaks around 5.5 \si{\electronvolt} damped slightly. On the other hand, for AuNP-HCQ complex after the adsorption of HCQ the plasmonic peak of gold in the range of 4.0 -- 4.5 \si{\electronvolt} exhibited the obvious variation in term of energy and intensity and shifting to blue, that established the more perturbative effect of HCQ adsorption on the electronic structure and plasmonic spectrum of gold nanoparticle in comparison to silver.
\\
\textbf{\large AgNP, AuNP, AgAuNP, and PtNP Coated with HCQ(CQ).} In this part the effect of changing the type of nanoparticles and increasing the number of HCQ molecules on the coating properties of nanoparticles are discussed based on the radial distribution function (RDF). RDF depicts how the density of one molecule changes as a function of distance from another reference molecule. In addition, the RDF can be used to represent the distance-dependent relative probability for observing a given site or atom relative to some central site or atom. This analysis provides the microstructure information about the arrangement of HCQ/CQ molecules as well as their affinity for interactions with nanoparticles\cite{brehm_travis_2011, ramezani_gold_2014, sambasivam_self-assembly_2016}. 

\begin{figure*}
%===============1st column RDF NPs - HCQ ======================
\begin{minipage}{.47\textwidth}  
	\begin{subfigure}{\linewidth}
	 \centering 
   	  \includegraphics[width=0.9\linewidth]{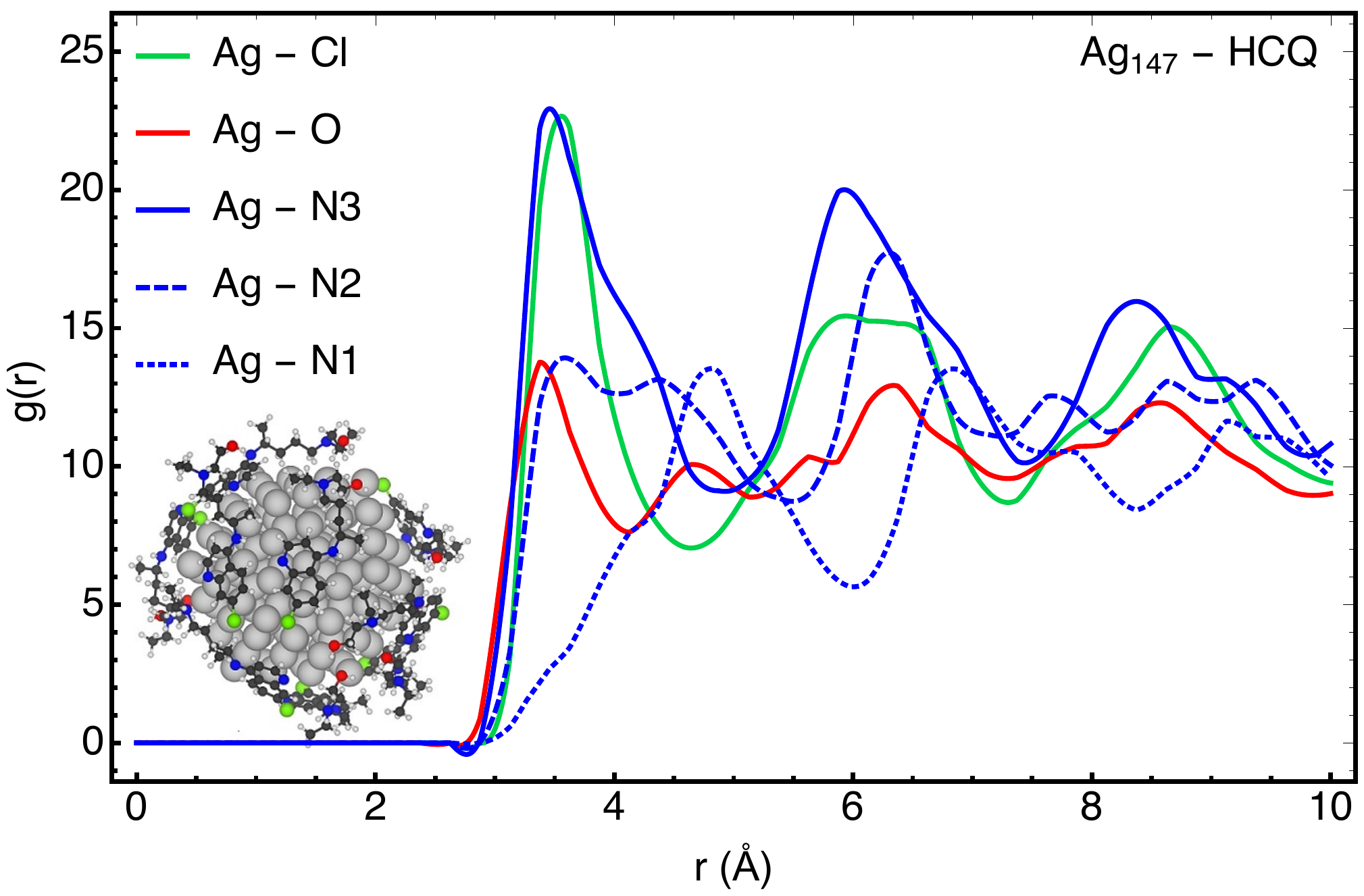}
	   \caption{}                                \label{fig_Atomic_RDF_AgHCQ} 
	  	 \end{subfigure}
\\
	\begin{subfigure}{\linewidth}
	  \centering  
 	    \includegraphics[width=0.9\linewidth]{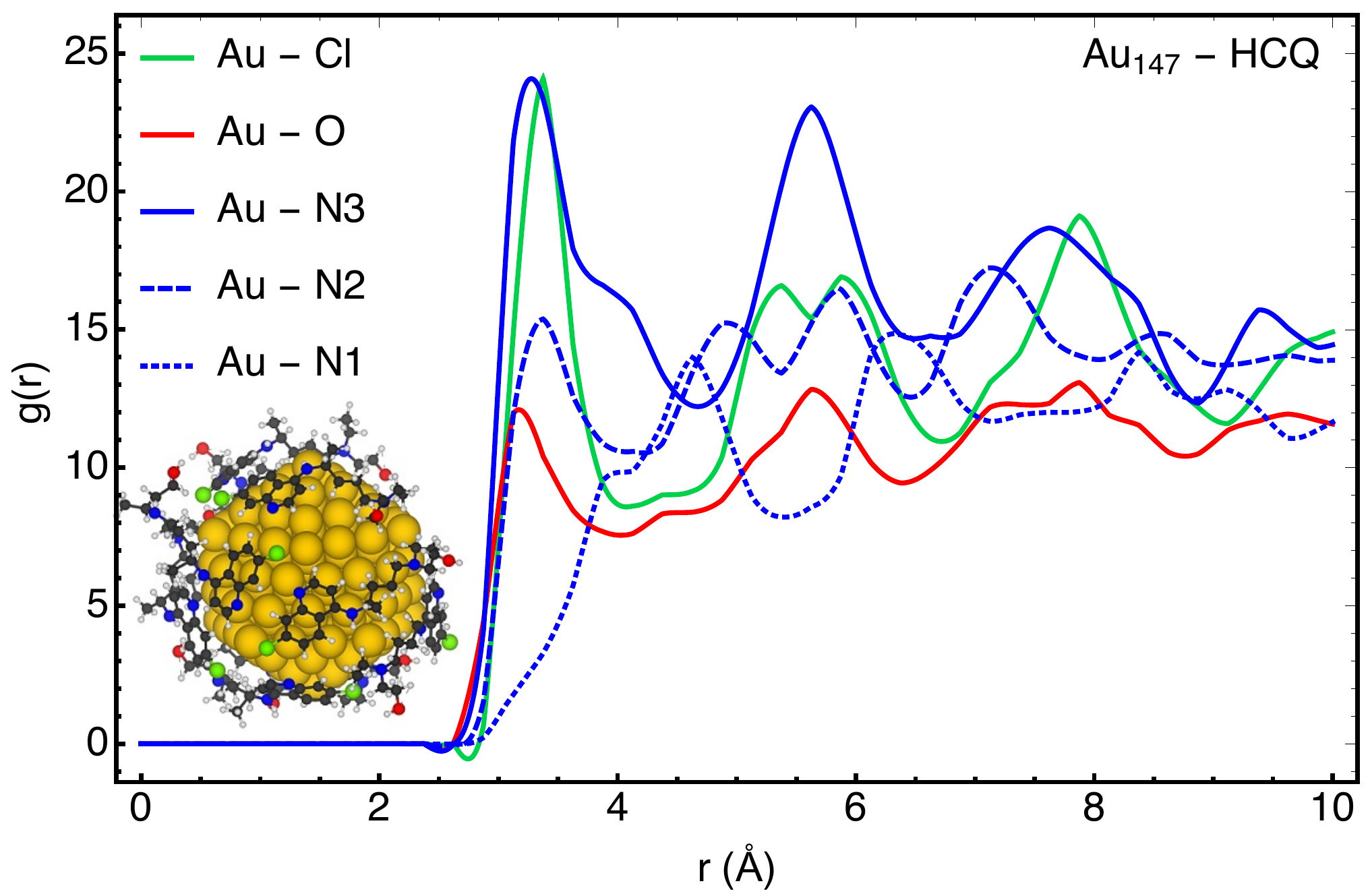}
   	  \caption{}                                \label{fig_Atomic_RDF_AuHCQ}
	\end{subfigure}
\\
\begin{subfigure}{\linewidth}
  \centering  
     \includegraphics[width=0.9\linewidth]{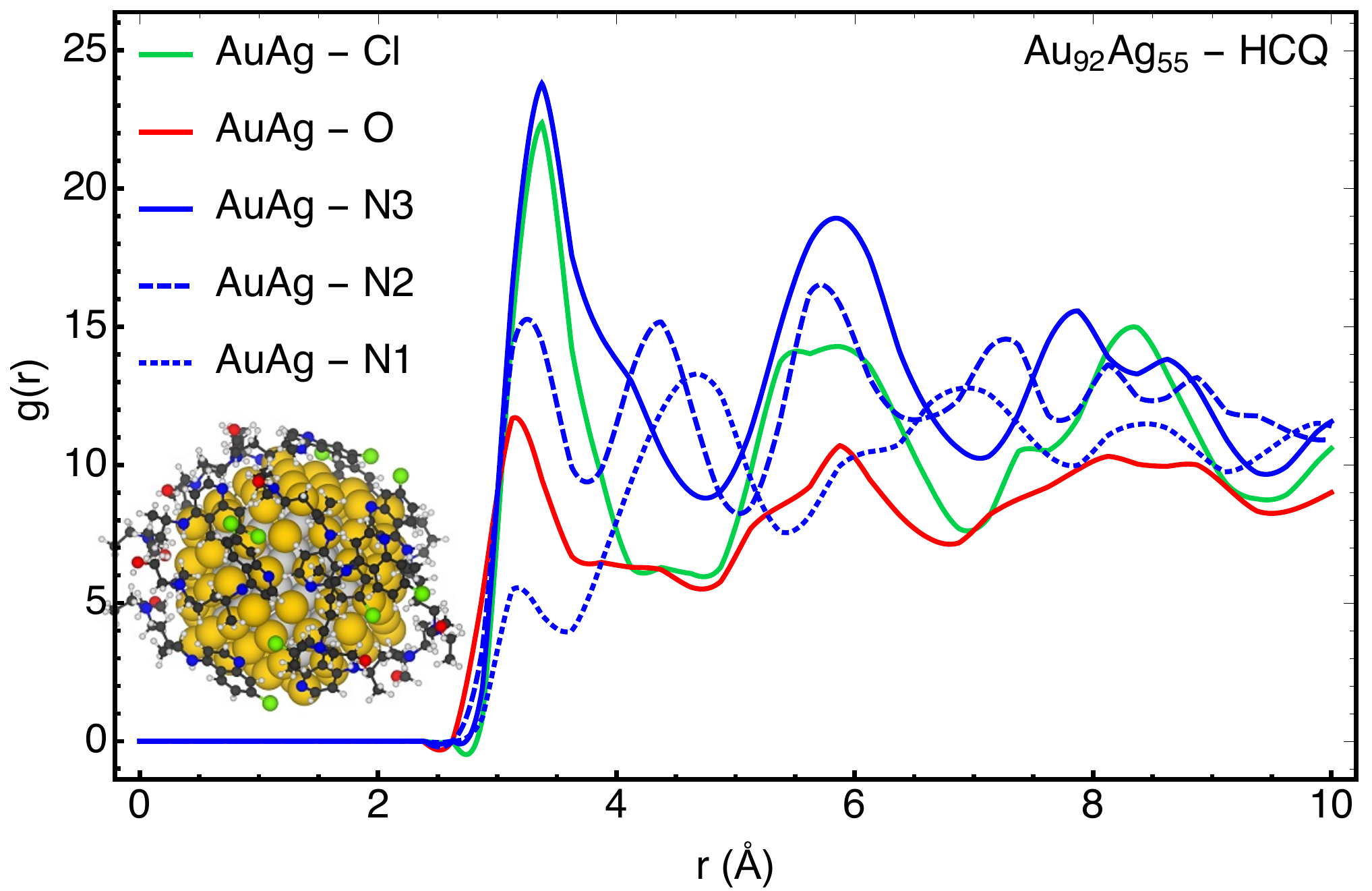}
        \caption{}                                \label{fig_Atomic_RDF_AuAgHCQ}
     \end{subfigure}
 \\
\begin{subfigure}{\linewidth}
  \centering  
     \includegraphics[width=0.9\linewidth]{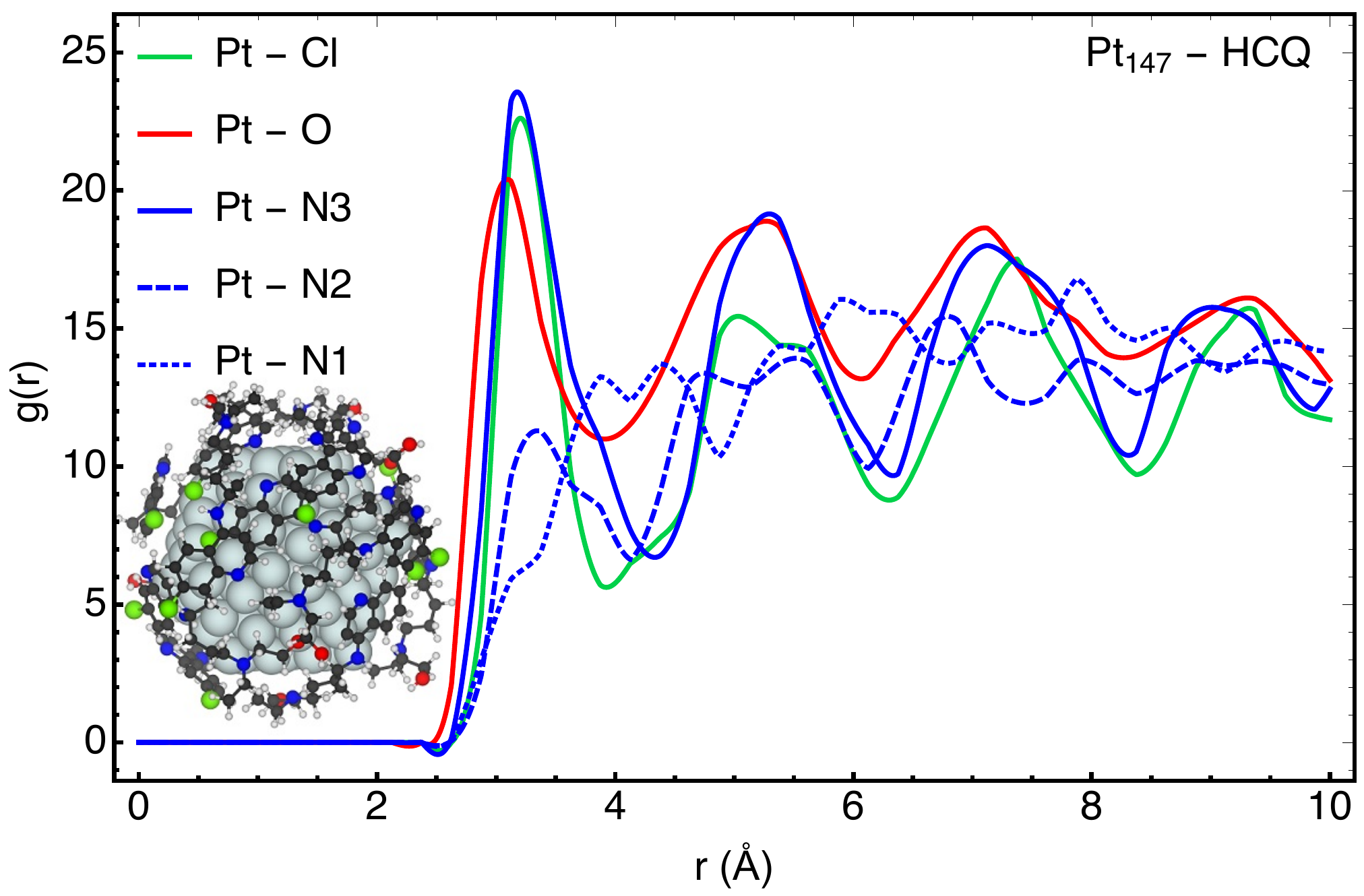}
     \caption{}                                \label{fig_Atomic_RDF_PtHCQ}
          \end{subfigure}
  \end{minipage}  
  %===============2nd column RDF NPs - CQ ======================
\begin{minipage}{.47\textwidth}  
\begin{subfigure}{\linewidth}
  \centering  
     \includegraphics[width=0.9\linewidth]{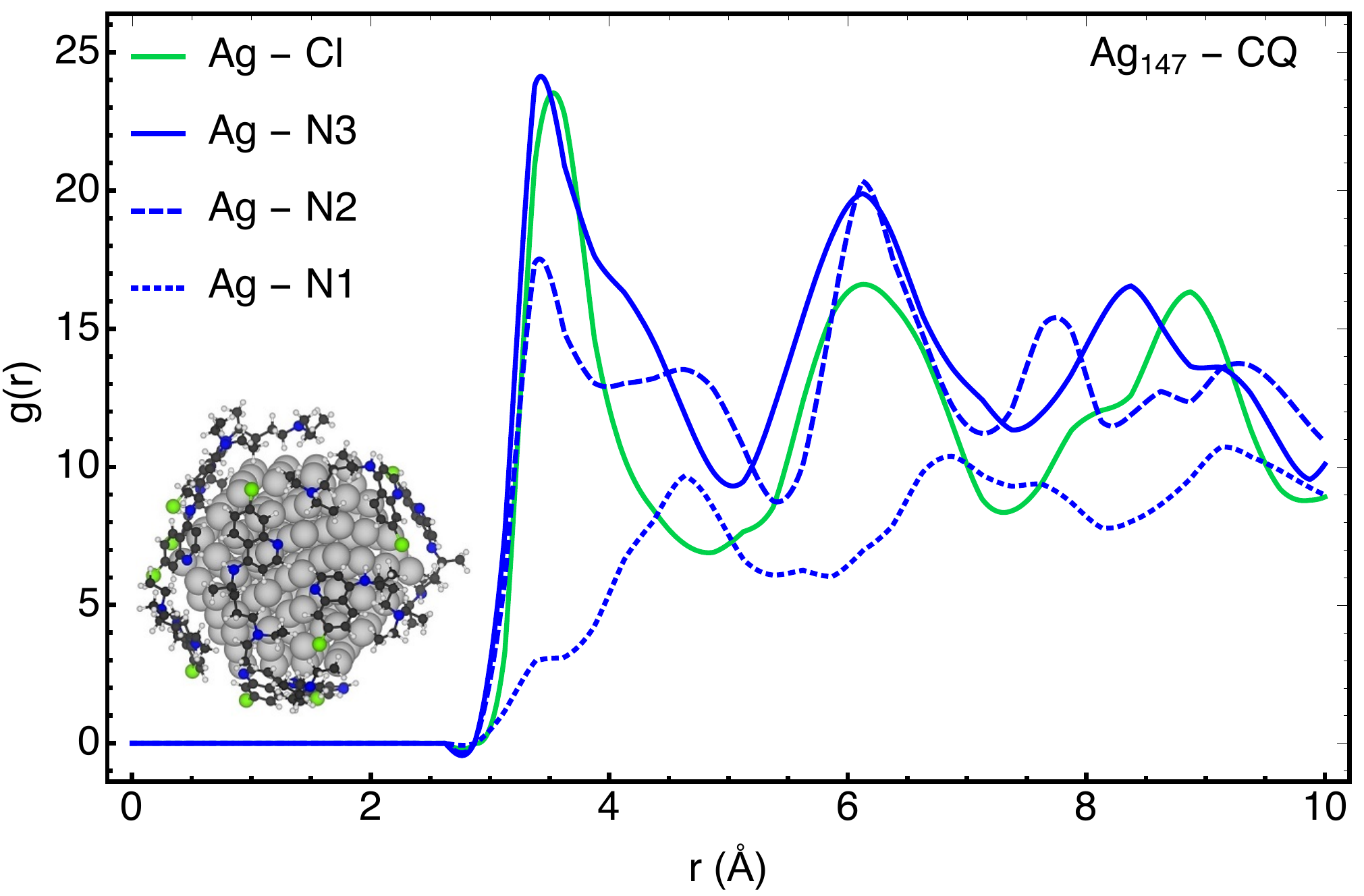}
      \caption{}                                \label{fig_Atomic_RDF_AgCQ}
     \end{subfigure}
\\
\begin{subfigure}{\linewidth}
  \centering  
     \includegraphics[width=0.9\linewidth]{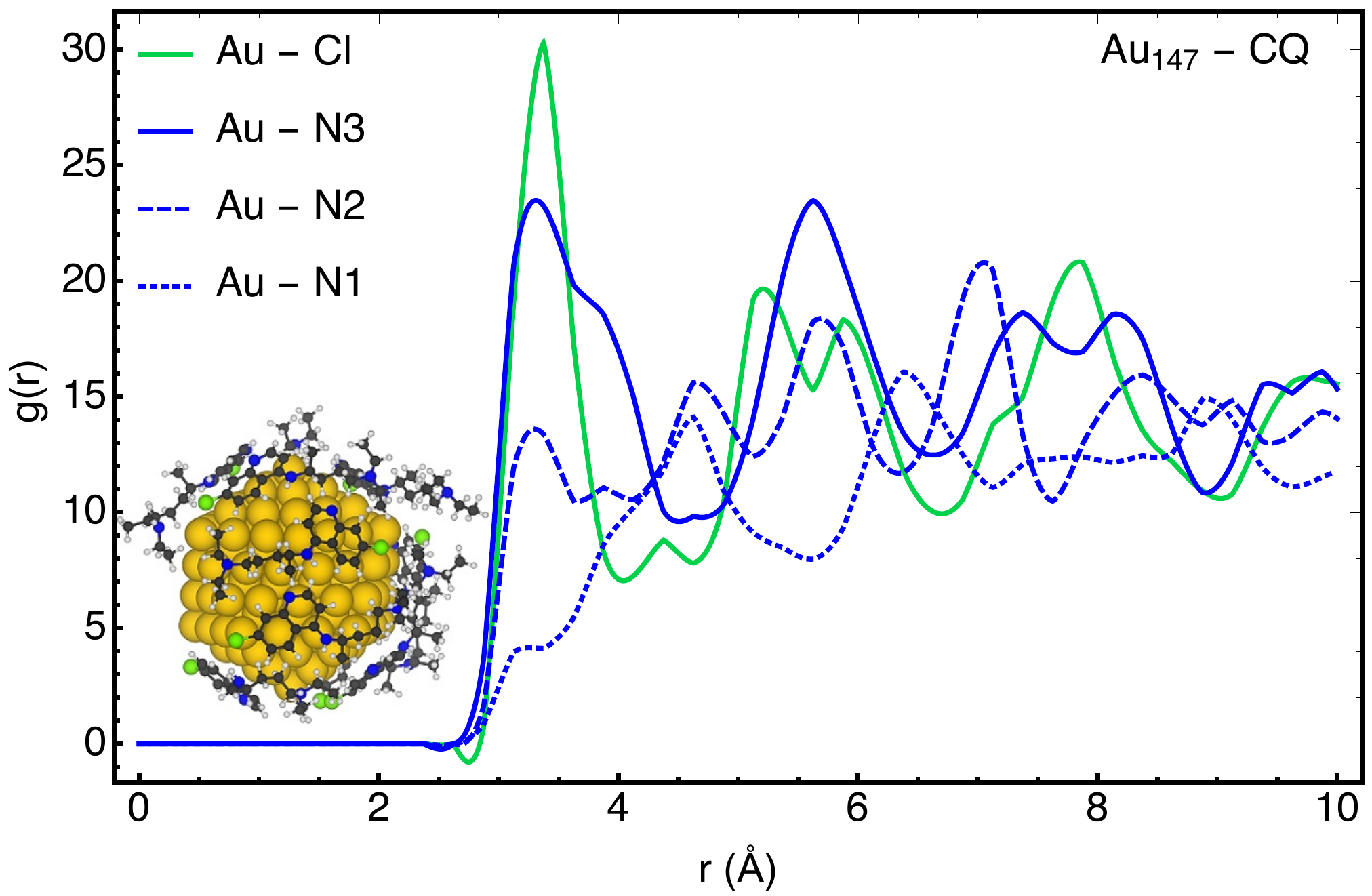}
     \caption{}                                \label{fig_Atomic_RDF_AuCQ}
     \end{subfigure}
\\
\begin{subfigure}{\linewidth}
  \centering  
     \includegraphics[width=0.9\linewidth]{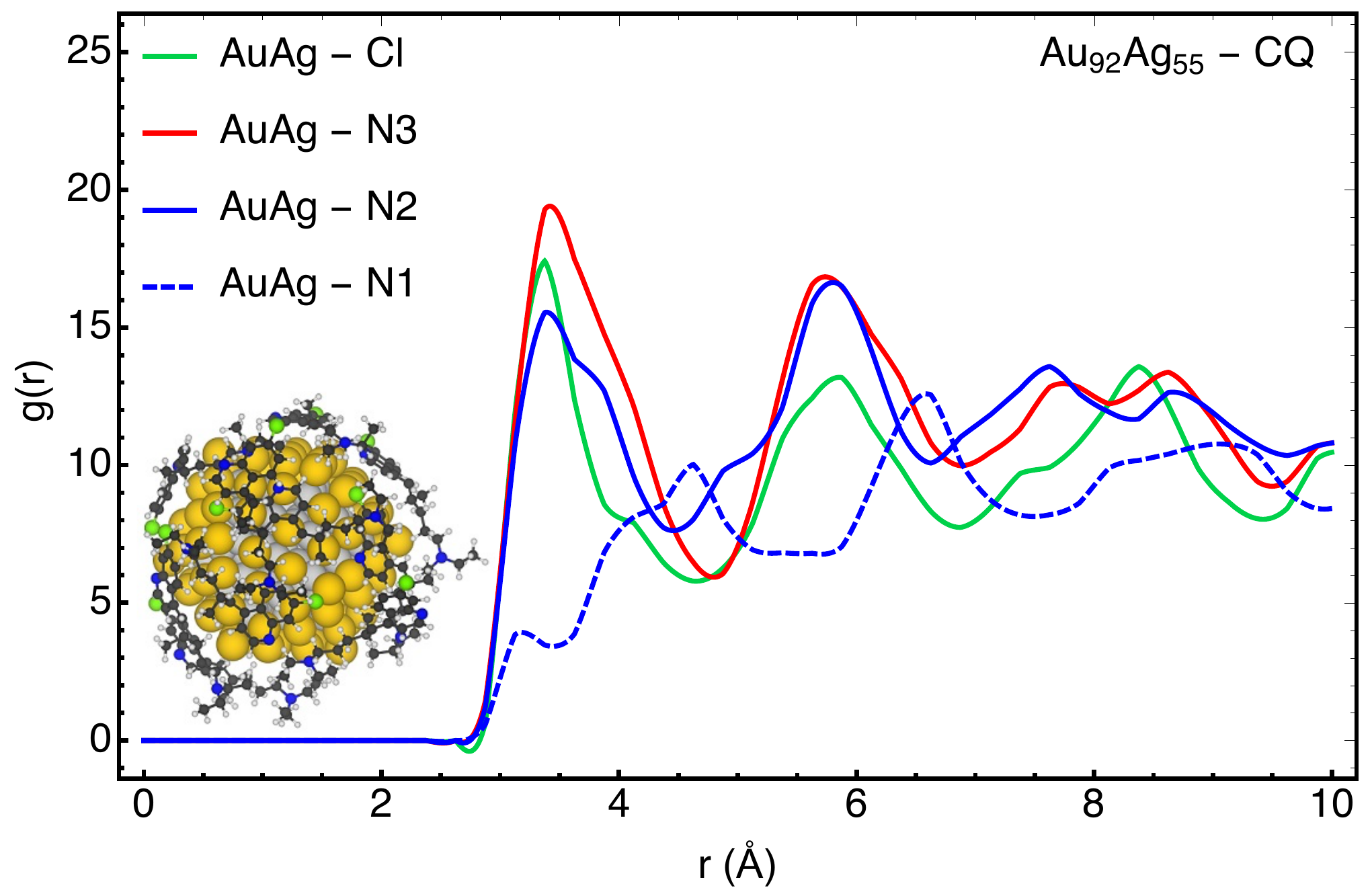}
     \caption{}                                \label{fig_Atomic_RDF_AuAgCQ}
     \end{subfigure}
 \\
\begin{subfigure}{\linewidth}
  \centering  
     \includegraphics[width=0.9\linewidth]{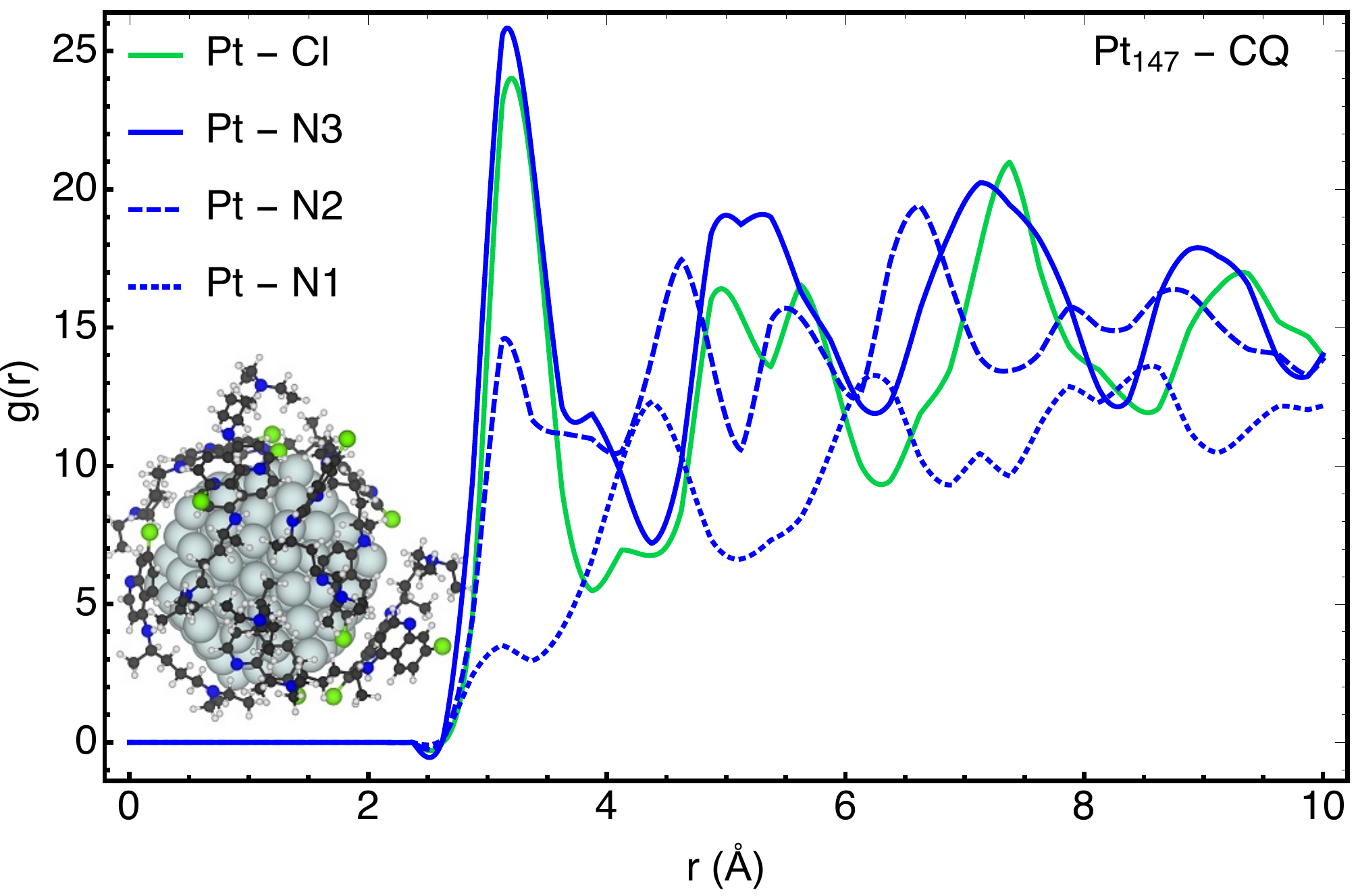}
     \caption{}                                \label{fig_Atomic_RDF_PtCQ}
     \end{subfigure}
  \end{minipage}
\caption{RDF plots for the active sites of HCQ/CQ with different type of (147- atomic icosahedral) nanoparticles such as AgNP, AuNP, AgAuNP, and PtNP. }
\label{fig_atomic_Nps_drugs}
\end{figure*}

\fig{fig_atomic_Nps_drugs} displays the RDF, g(r), for the active sites of HCQ/CQ molecules with different type of (147-atomic icosahedral) nanoparticles such as AgNP, AuNP, AgAuNP, and PtNP. As can be found in \fig{fig_atomic_Nps_drugs}, the nitrogen of pyridine ring has the highest affinity in comparison to other type of nitrogen that is in agreement with DFT calculation (in the previous section). However, it seems in \fig{fig_atomic_Nps_drugs}, the peak of chlorine grows up same as nitrogen, that can be related to its vicinity to nitrogen atom of pyridine ring and not the affinity of Cl-group. The optimized structure of DFT and the lower negative charge of Cl (-0.03 $|e|$) in comparison to N (-0.17 $|e|$) are the acceptable reasons for this claim. In addition, the coating of \ce{Ag147} with more HCQ molecules have simulated (in Figure S5) and confirmed the higher g(r) for N- respect to Cl-group. Furthermore, the O-atom of hydroxyl group is another active site of HCQ for interaction, however, for gold, silver and alloy nanoparticle, the g(r) values for O-group is lower than N-group and for PtNP, its g(r) is slightly more than N-group. The more attraction of N-atom (verse O-group) for interaction with gold and silver noble metal was established by A. Antusek et al based on ab-initio calculation\cite{antusek_lone_2003}. The probability distribution map of atoms near the nanoparticles (see Figure S2–S3 in supporting information), has confirmed the RDF results. 

\begin{figure*}
%=============   NPs - atoms   ==========================
\begin{subfigure}{.32\textwidth}
  \centering
  \includegraphics[width=\linewidth]{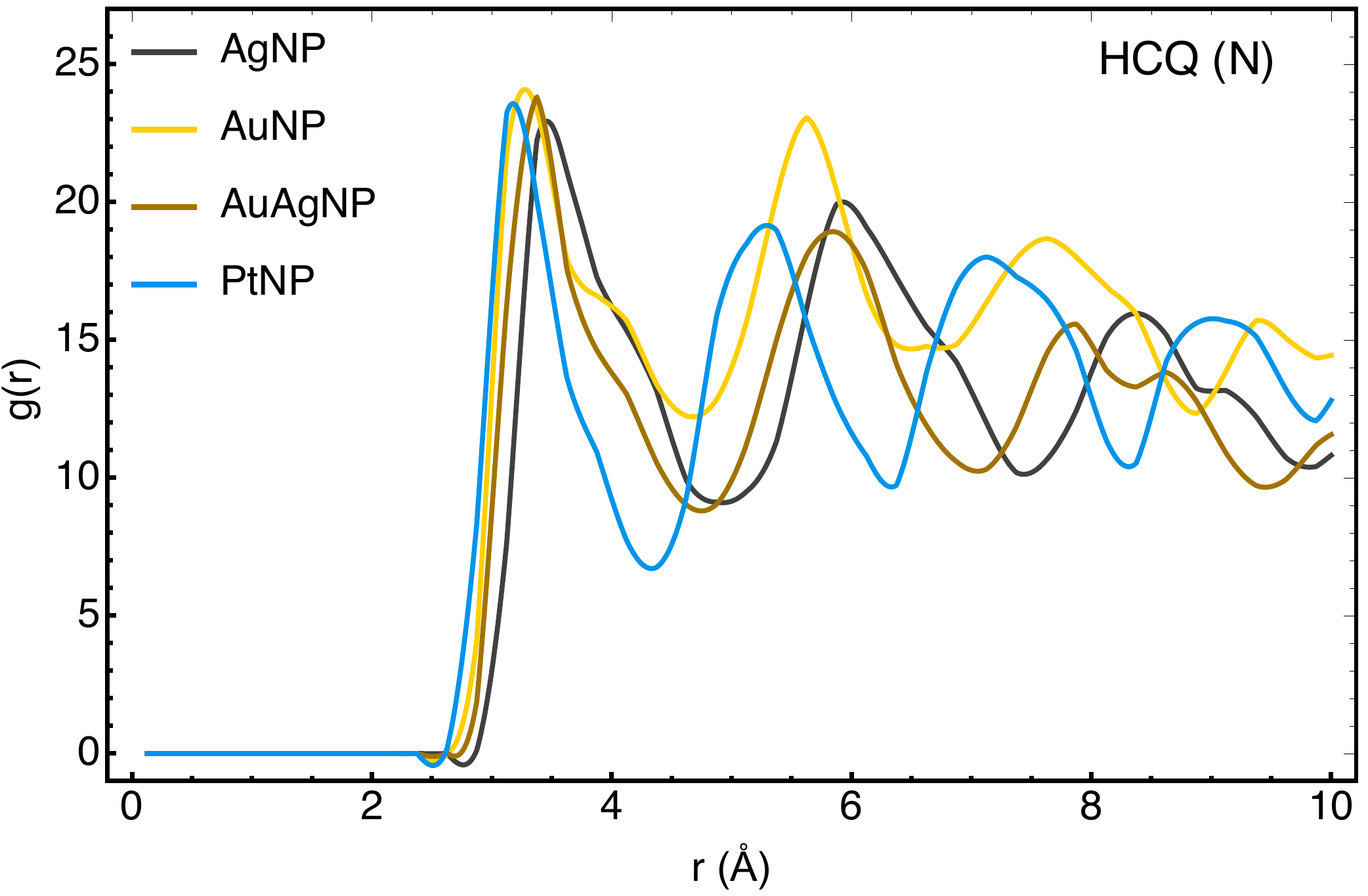}
     \caption{}
  \label{fig_NPs-N(HCQ)}
\end{subfigure}
%_________________________
\begin{subfigure}{.32\textwidth}
  \centering
  \includegraphics[width=\linewidth]{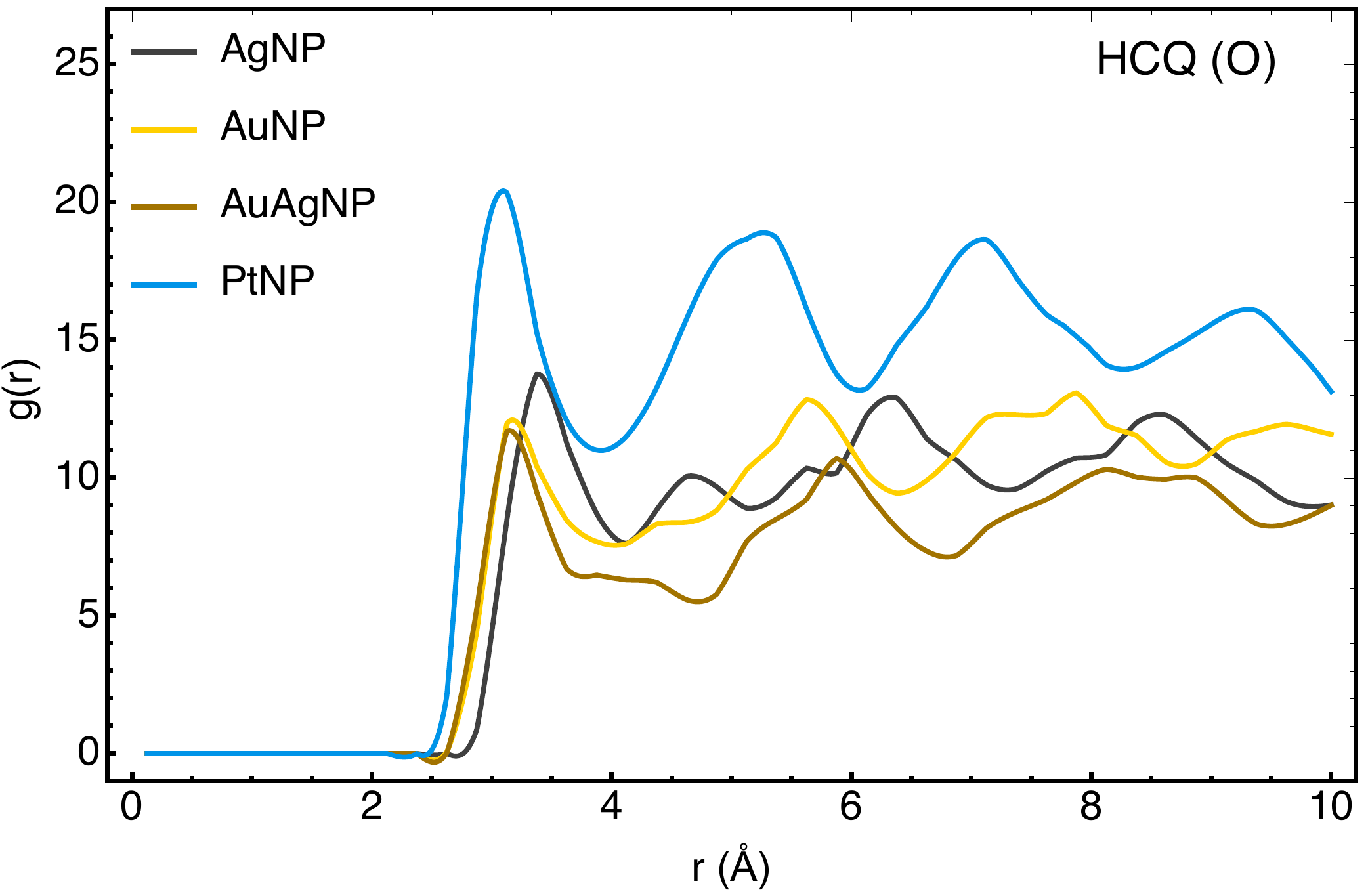}
   \caption{}
  \label{fig_NPs-O(HCQ)}
\end{subfigure}
%_________________________
\begin{subfigure}{.32\textwidth}
  \centering
  \includegraphics[width=\linewidth]{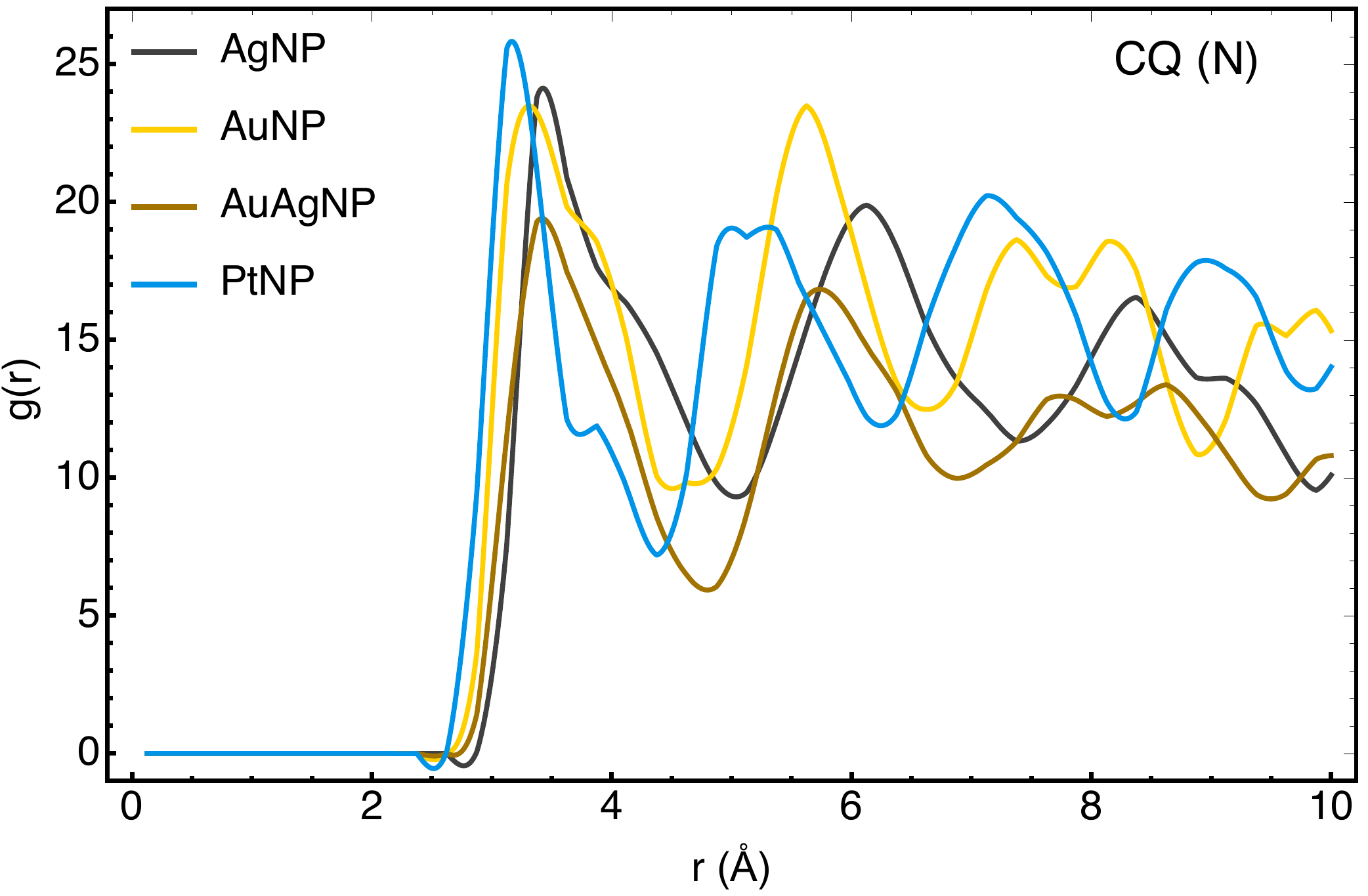}
     \caption{}
  \label{fig_NPs-N(CQ)}
 \end{subfigure}
  %=============   NPs - HCQ_vs_CQ   ==========================
\begin{subfigure}{.32\textwidth}
  \centering
  \includegraphics[width=\linewidth]{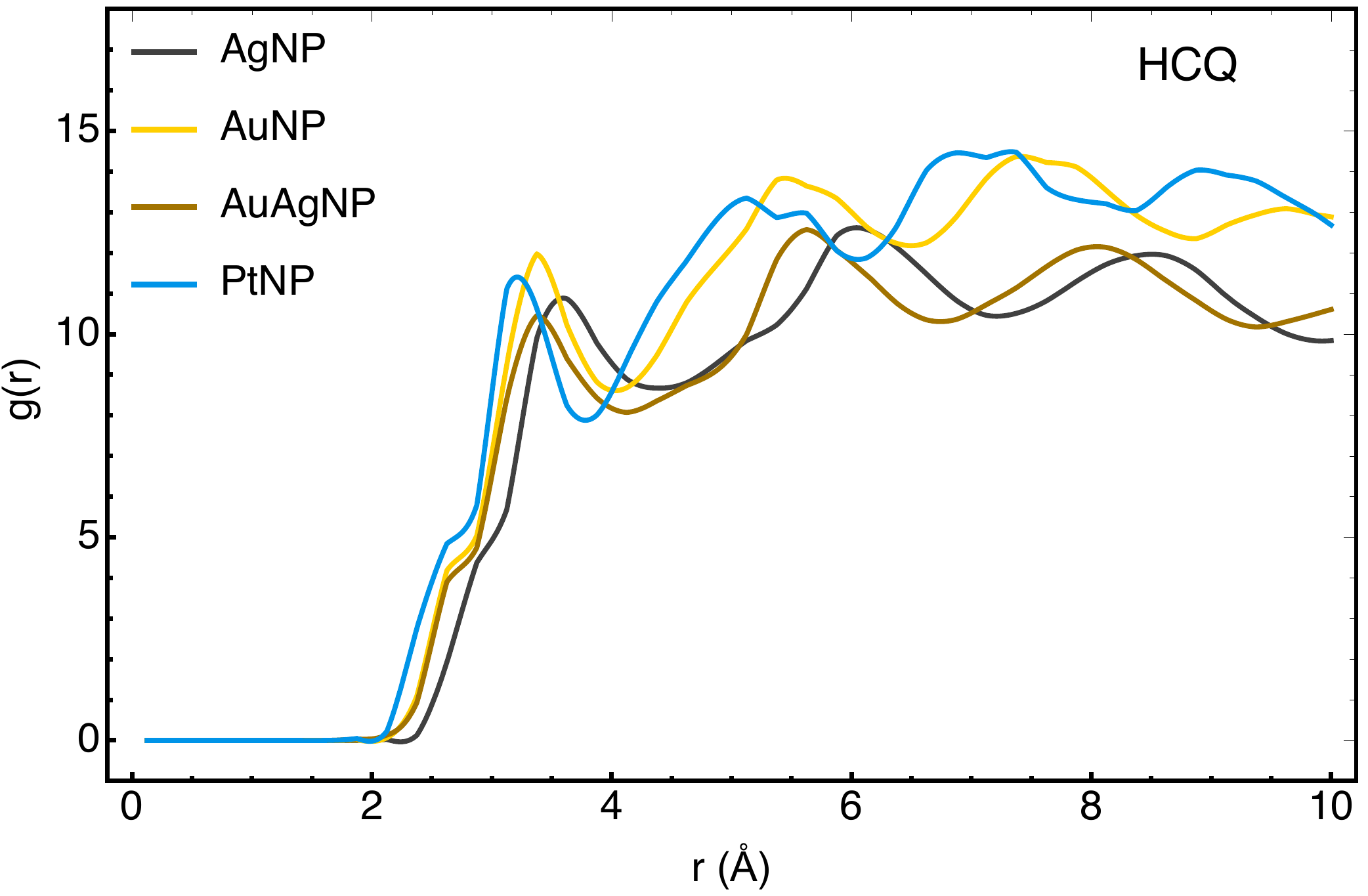}
   \caption{}
  \label{fig_NPs-HCQ}
\end{subfigure}
%_________________________
\begin{subfigure}{.32\textwidth}
  \centering
  \includegraphics[width=\linewidth]{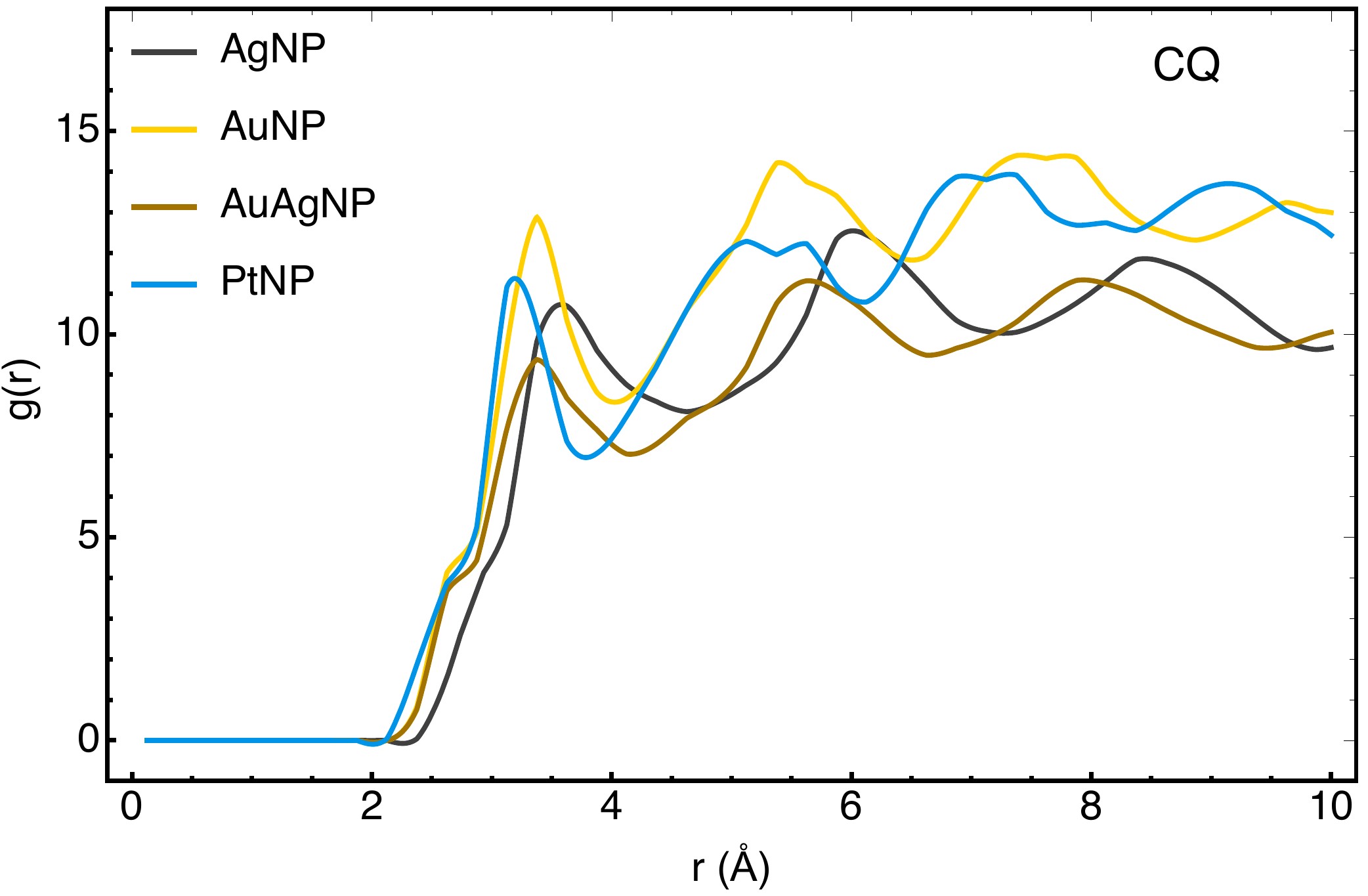}
   \caption{}
  \label{fig_NPs-CQ}
\end{subfigure}
%_________________________
\begin{subfigure}{.32\textwidth}
  \centering
  \includegraphics[width=\linewidth]{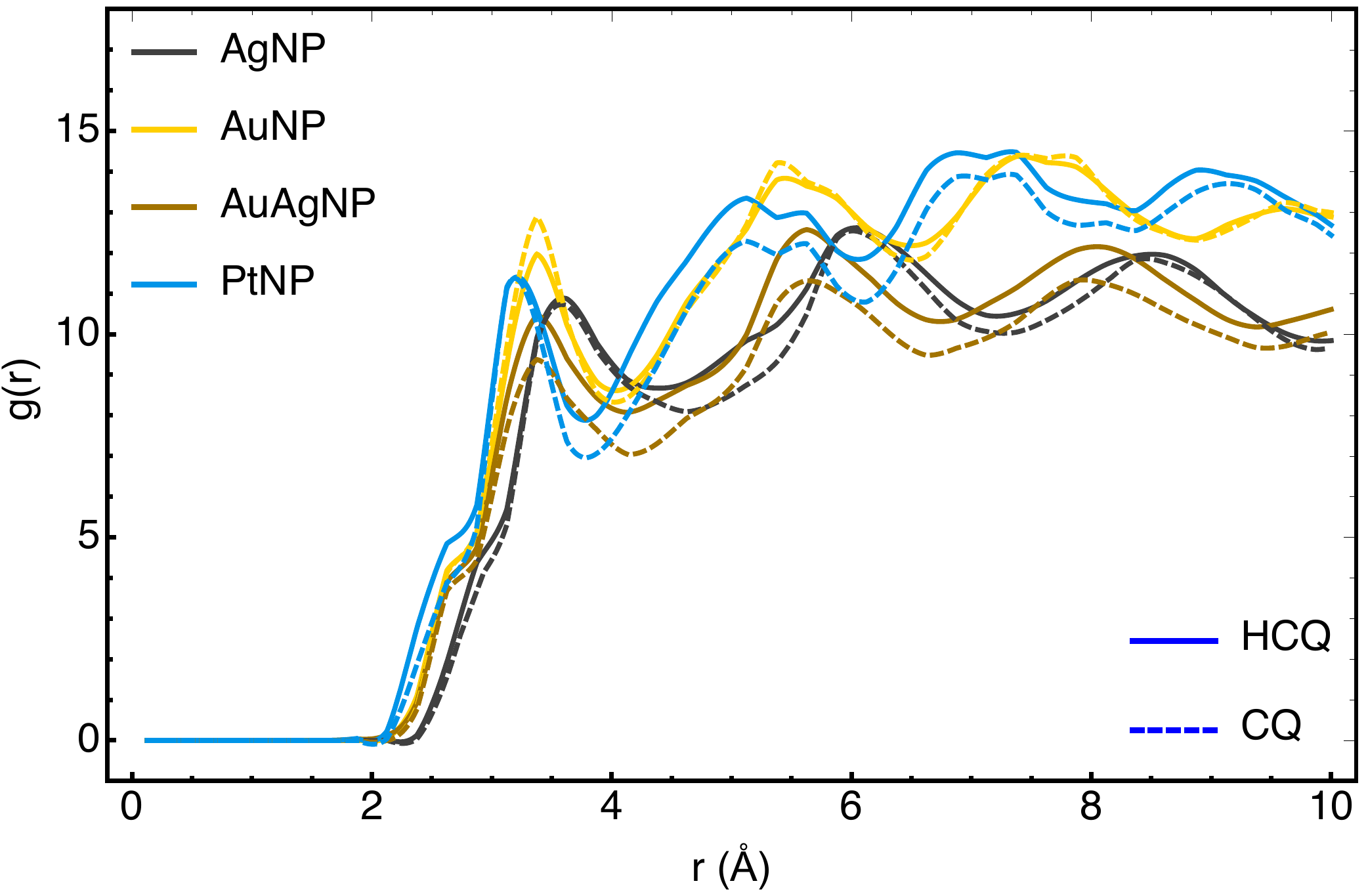}
   \caption{}
  \label{fig_Nps_HCQ_vs_CQ}
\end{subfigure}
\caption{Comparison the total and atom type RDF plots for HCQ/CQ molecules with AgNP, AuNP, AgAuNP, and PtNP.}
\label{fig_Nps_drugs}
\end{figure*}

\fig{fig_Nps_drugs} compares the total and atom type RDFs of HCQ/CQ molecules with respect to the variation of nanoparticles to propose the possible candidates for adsorption. \figtwo{fig_Nps_drugs}{fig_NPs-N(HCQ)}, \figtwo{fig_Nps_drugs}{fig_NPs-O(HCQ)}, and \figtwo{fig_Nps_drugs}{fig_NPs-N(CQ)} compare the RDF of N- and O-atom of HCQ and N-atom of CQ with respect to the type of nanoparticles. For N-group the affinity of different type of nanoparticles is nearly similar to each other, although PtNP has strong numerous peaks in the range of $<$ 10 \si{\angstrom}. On the other hand, the sharp and intense g(r) peak for O-group with PtNP which is appeared in the shorter distance demonstrates the best adsorption properties on PtNP and also explains the stronger affinity of PtNP to HCQ compare to  CQ. Accordingly, we can conclude that HCQ preferred to adsorb on AuNP, AgNP and alloy from one side and coated on PtNP by N- and O-group. 

Finally, \figtwo{fig_Nps_drugs}{fig_NPs-HCQ} and \figtwo{fig_Nps_drugs}{fig_NPs-CQ} show the total RDF plots of HCQ and CQ molecules with respect to type of nanoparticles, respectively and represents the overall coating trend as follow: PtNP $>$ AuNP $>$ AuAgNP $>$ AgNP. In addition, comparing the total RDF of NPs with HCQ and CQ (\figtwo{fig_Nps_drugs}{fig_Nps_HCQ_vs_CQ}) indicates that PtNP has a bigger g(r) for HCQ compare to CQ which is in agreement with our DFT results presented in Table S8. 
\\
\\
\textbf{\large Coated HCQ on Different Size of AgnNP (n = 147, 561, 1415, and 2869).} The effect of size on the adsorption properties of AgNP has investigated for the fixed number of HCQ (12 molecules) that interacted with twelve active sites in the corners of icosahedral AgNP and also different number of HCQ that increased (to 32, 64, 105 molecules) proportional to the numbers of AgNP surface atoms. The radius of \ce{Ag147}, \ce{Ag561}, \ce{Ag1415} and \ce{Ag2869} nanoparticles is calculated to be 1.6, 2.6, 3.6, and 4.6, \si{\nano\metre} respectively while their thickness after coating with HCQ has been increased about \SI{1}{\nano\metre}. 

Figure S4 shows the coated nanoparticles with HCQ molecules exhibited the significant fluctuation within at the beginning of the simulations ($\sim$ 5 \si{\nano\second} ), indicating HCQ molecules are moving free near nanoparticles due to spatial setting of the medicines in the active site. After the fluctuation, it maintained a continuous equilibrium up to the end of the simulation time. These results suggest that the stabilities of the dynamic equilibriums for the complexes were reliable and that the trajectories could be useful in collecting snapshots for further analyses (see Figure S4).

\begin{figure*}
%===============1st column  : RDF Agx-HCQ12======================
\begin{minipage}{.36\textwidth}  
\begin{subfigure}{\linewidth}
  \centering  
     \includegraphics[width=\linewidth]{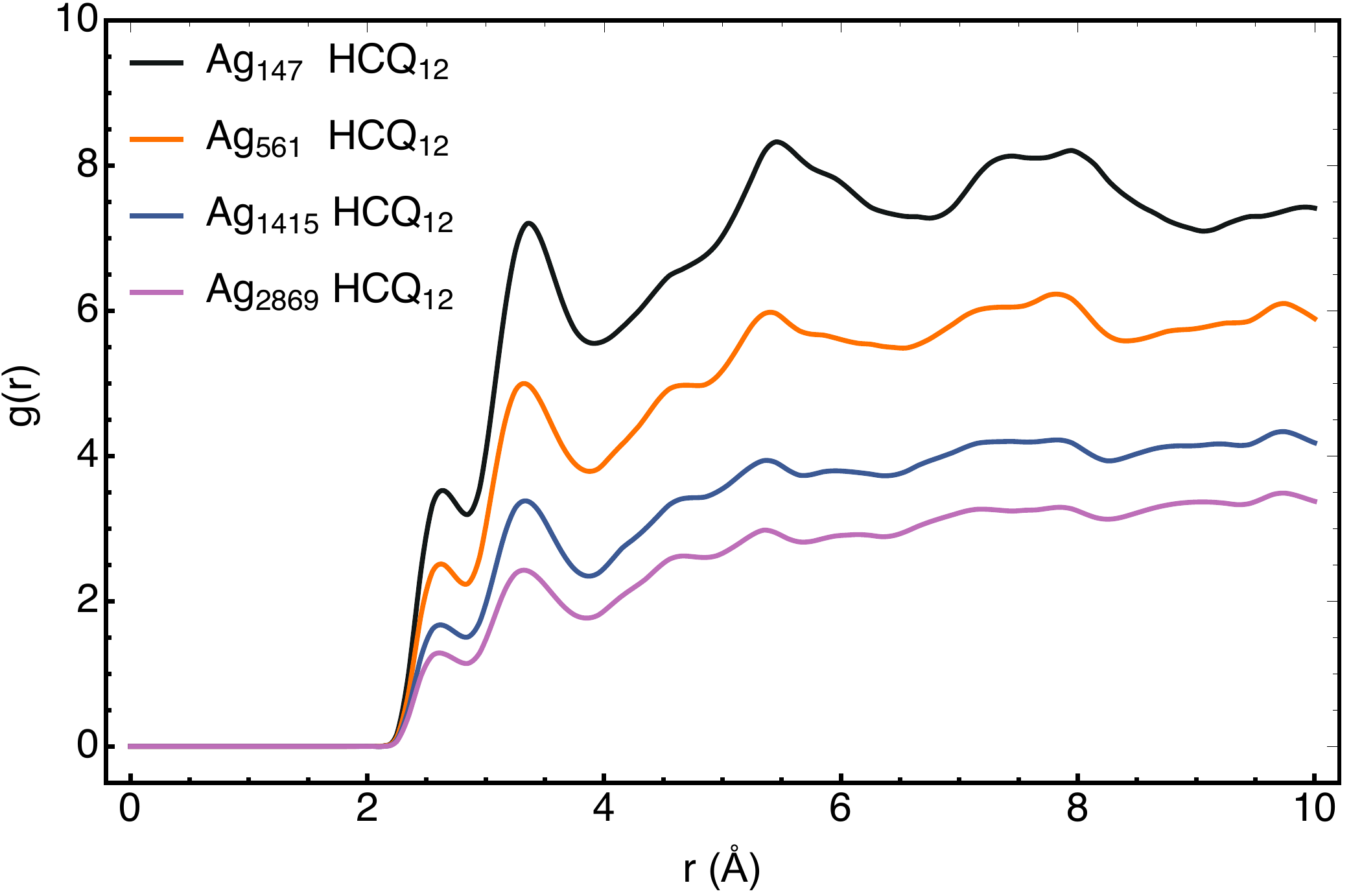}
     \caption{}                                                        \label{fig_RDF_Agx-HCQ12}
     \end{subfigure}
\\
 \begin{subfigure}{\linewidth}
  \centering  
  \includegraphics[width=\linewidth]{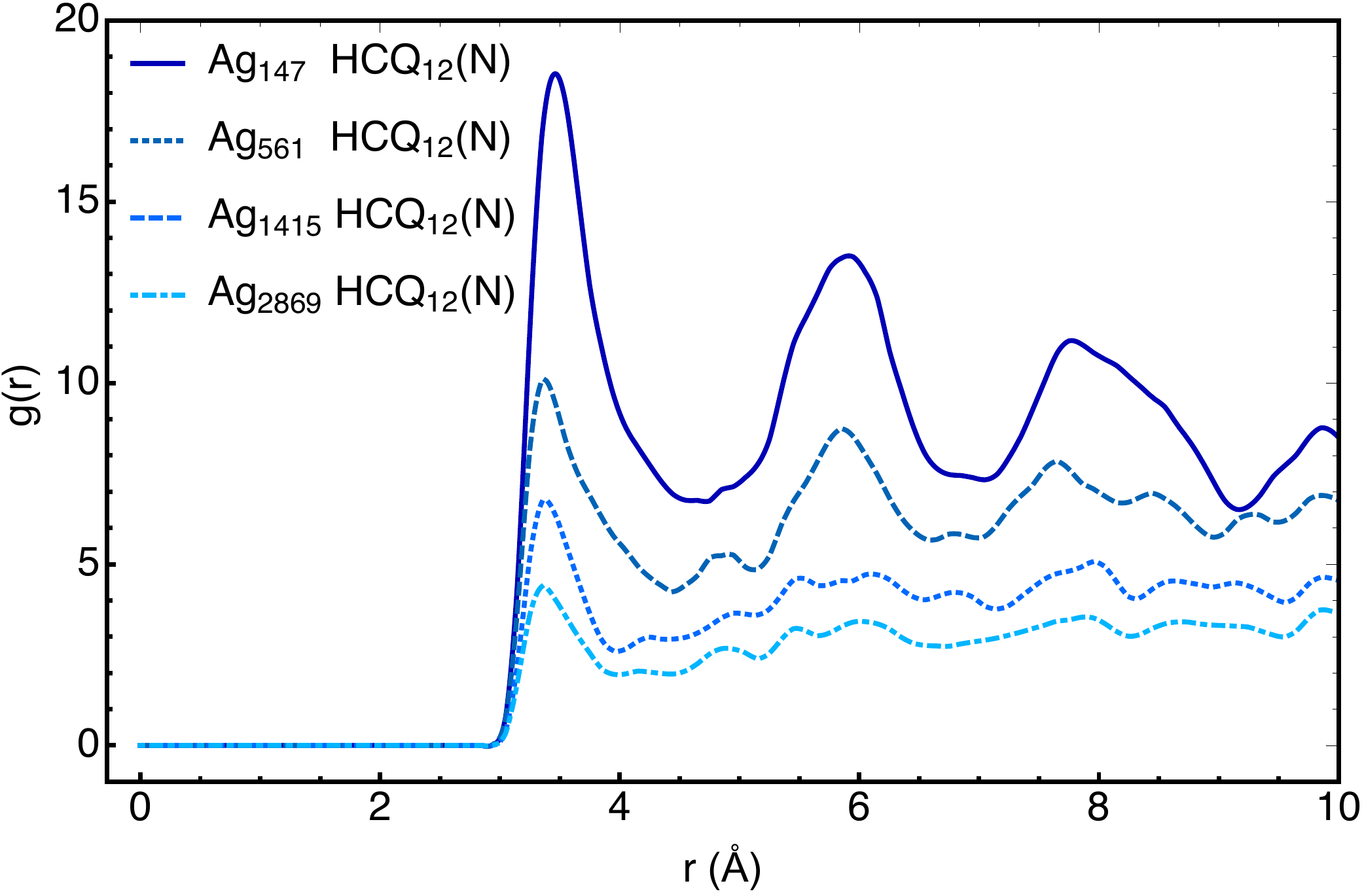}
    \caption{}  							\label{fig_RDF_Agx-N(HCQ12)}
\end{subfigure}
  \end{minipage}
%=================2nd column  : NPs_HCQ12==========  
\begin{minipage}{.12\textwidth}  \center
  \begin{subfigure}{\linewidth}
   \centering
     \includegraphics[width=1.1\linewidth]{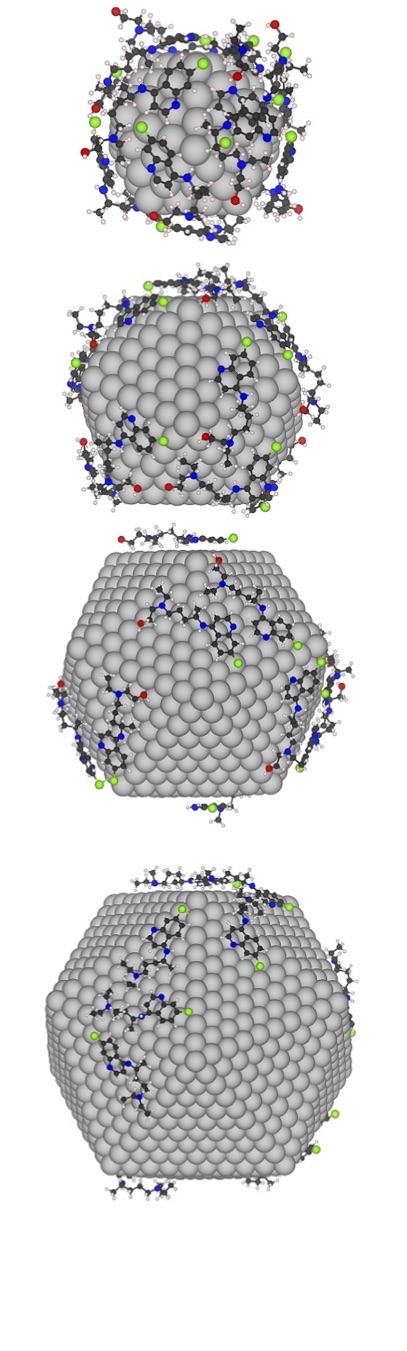}
   \end{subfigure}
      \end{minipage}
%===============3rd column  : Agx-HCQy=======,height=0.1\linewidth===============
\begin{minipage}{.36\textwidth}  
\begin{subfigure}{\linewidth}
  \centering  
     \includegraphics[width=\linewidth]{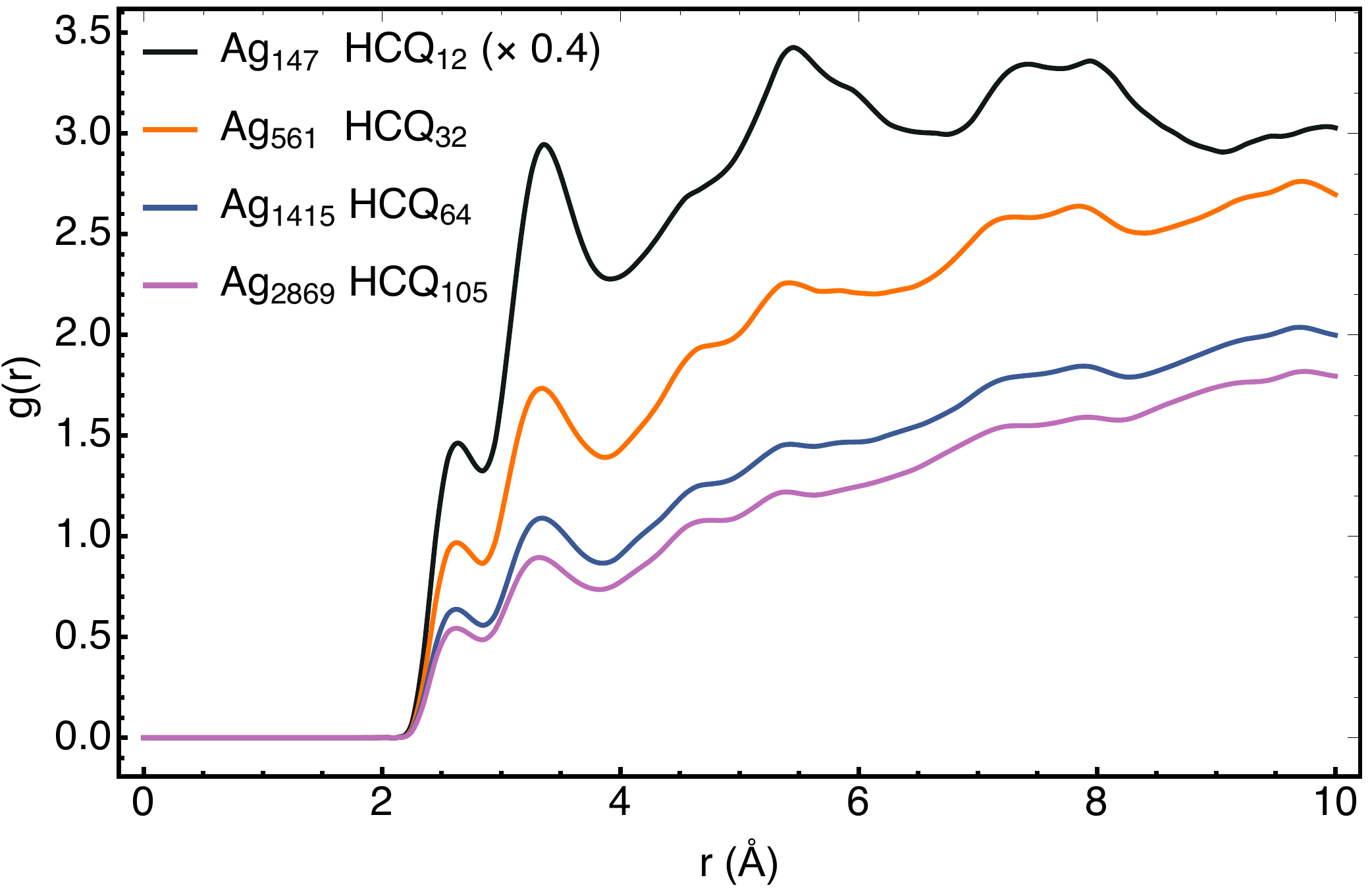}
     \caption{}                                                        \label{fig_RDF_Agx-HCQy}
     \end{subfigure}
\\
 \begin{subfigure}{\linewidth}
  \centering  
  \includegraphics[width=\linewidth]{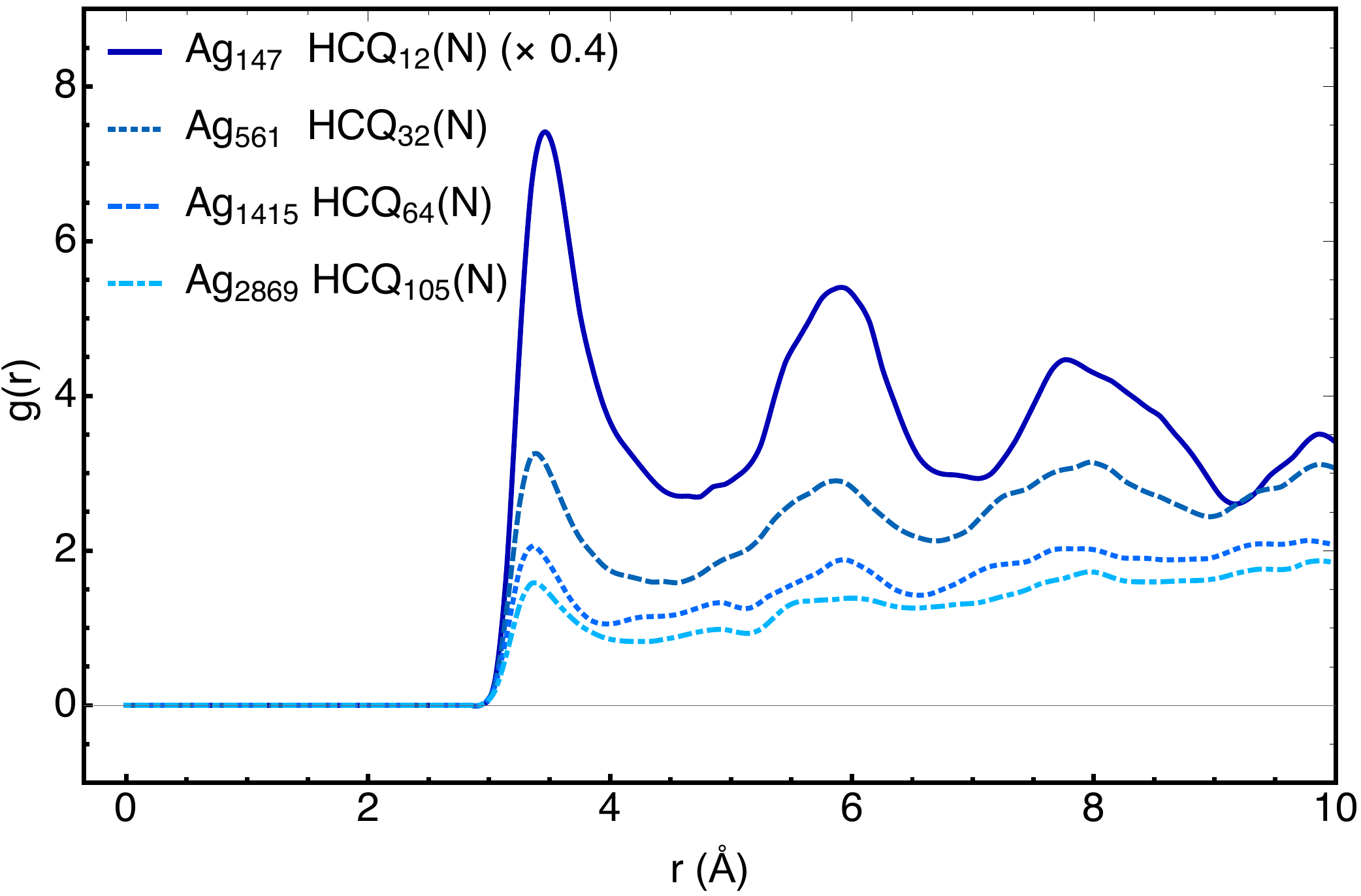}
    \caption{}  							\label{fig_RDF_Agx-N(HCQy)}
\end{subfigure}
  \end{minipage}
  %=================2nd column  : NPs_HCQ y==========  
\begin{minipage}{.12\textwidth}  \center
     \includegraphics[width=1.1\textwidth]{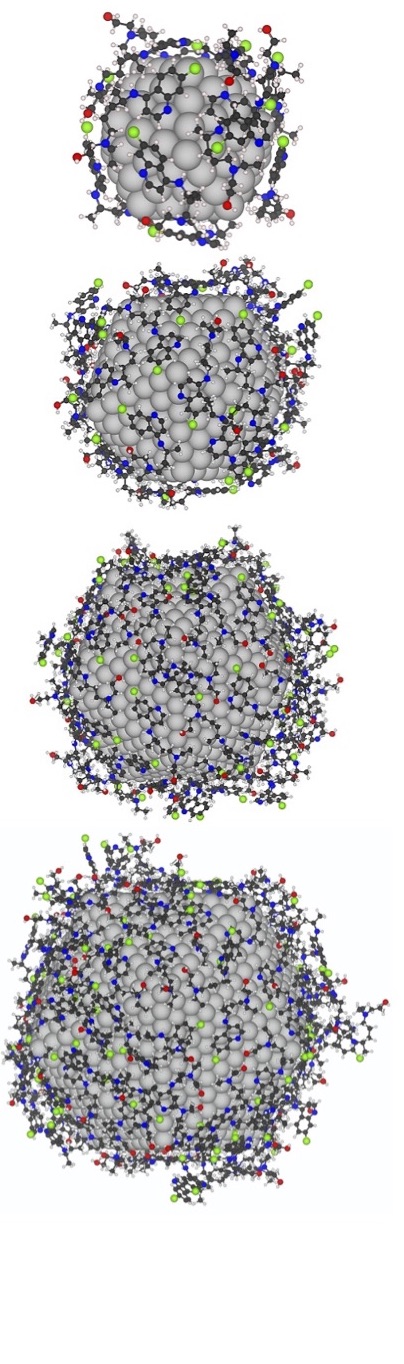}
  \end{minipage}
%=======================================
\caption{RDF of HCQ with different size of AgNPs. (\subref{fig_RDF_Agx-HCQ12}),(\subref{fig_RDF_Agx-N(HCQ12)}) AgNPs are coated with the same number of HCQ molecules. (\subref{fig_RDF_Agx-HCQy}), (\subref{fig_RDF_Agx-N(HCQy)}) AgNPs are coated with different number of HCQ molecules}
\label{fig_RDF_Npx_HCQy}
\end{figure*}

In \fig{fig_RDF_Npx_HCQy} (and Figure S5 and S6) the RDF plots for the fixed and different number of HCQ with respect to the size of AgNPs and type of anchoring atoms are displayed. The comparison of RDF plots for the N-group as an active-site (\figtwo{fig_RDF_Npx_HCQy}{fig_RDF_Agx-N(HCQ12)}), in addition to all atoms (\figtwo{fig_RDF_Npx_HCQy}{fig_RDF_Agx-HCQ12}) exhibited the decreasing in the adsorption properties by going from small to larger AgNPs. The same trend was reproduced in \figtwo{fig_RDF_Npx_HCQy}{fig_RDF_Agx-HCQy} and \figtwo{fig_RDF_Npx_HCQy}{fig_RDF_Agx-N(HCQy)} in which the portion of HCQs increased with respect to the number of surface's atoms. Generally, the overall coating properties decreases about 65-85\% by increasing the size of nanoparticle from 1.6  to 4.6 \si{\nano\metre}, see Figure S7. It has been established that the affinity of noble metal clusters enhanced by decreasing the size.  Moreover, in agreement with the previous section the N-group has the higher affinity for interaction with different size of AgNPs with respect to O-group (see Figure S5 and S6).  Finally, Figure S8 shows the RDF comparison for HCQ and CQ coated on \ce{Ag2869}, and reveals the appropriate adsorption affinity of HCQ versus CQ. 

\section{Conclusion}
In summary, the adsorption properties of noble metal nanoparticles and their coating with HCQ/CQ as the promising candidate for COVID-19 treatment have been studied. The weak charge-transfer interaction with partially negative charge N- and O-group of drugs that increased by changing the type of metal nanoparticles element  (PtNP $>$ AuNP $>$ AuAgNP $>$ AgNP) have been investigated. Finally, this kind of combining the quantum mechanics and molecular dynamic simulation can be suggesting these noble nanoparticles (with low toxicity and antiviral activity) as the vehicle for efficient usage of HCQ/CQ and decreased the drugs side effect.    

\section*{Acknowledgment}
The authors acknowledge the UNESCO UNISA ITHEMBA-LABS/NRF Africa Chair in Nanosciences \& Nanotechnology (U2ACN2), the Centre for High Performance Computing (CHPC), South Africa, and High Performance Computing Centre (HPCC) of Sharif University of Technology, Iran, for providing computational resources and facilities for this research project. Z. J. acknowledges the developer group of Software for Chemistry \& Materials (SCM) and computing resources of VU University of Amsterdam.

\begin{suppinfo}
Table S1--S6: Optimized geometry data of CQ, HCQ, \ce{Ag147}, \ce{Au147}, \ce{Au92Ag55}, and \ce{Pt147} from Density Functional Theory. 
Table S7: The Lennard-Jones parameters for nanoparticles simulation. 
Table S8: The interaction energy of HCQ and CQ molecules with Ag (111), Au (111), and Pt (111) surfaces.
Figure S1: Charge distribution of HCQ and CQ molecules and their electrostatic potential map.
Figure S2: Probability distribution of Cl, N (in pyridine ring) and O of HCQ near the \ce{Ag147}, \ce{Ag561}, \ce{Ag1415} and \ce{Ag2869}.
Figure S3: Probability distribution of Cl and N (in pyridine ring) of CQ near the \ce{Ag147}, \ce{Ag561}, \ce{Ag1415} and \ce{Ag2869}.
Figure S4: RMSD of different size of AgNP coated with HCQs.
Figure S5: The comparison of RDFs for \ce{Ag147}, \ce{Ag561}, \ce{Ag1415} and \ce{Ag2869}  with 12 molecules of HCQ.
Figure S6: The comparison of RDFs for \ce{Ag147}, \ce{Ag561}, \ce{Ag1415} and \ce{Ag2869} with 12, 32, 64, and 105 molecules of HCQ, respectively.
Figure S7: Decreasing the overall coating properties by increasing the size of AgNP from 1.6  (\ce{Ag147}) to 4.6 (\ce{Ag2869}) \SI{4.6}{\nano\meter}.
Figure S8: The comparison of RDFs for HCQ and CQ with \ce{Ag2869}.

\end{suppinfo}

%\bibliography{refs}
\providecommand{\latin}[1]{#1}
\makeatletter
\providecommand{\doi}
  {\begingroup\let\do\@makeother\dospecials
  \catcode`\{=1 \catcode`\}=2 \doi@aux}
\providecommand{\doi@aux}[1]{\endgroup\texttt{#1}}
\makeatother
\providecommand*\mcitethebibliography{\thebibliography}
\csname @ifundefined\endcsname{endmcitethebibliography}
  {\let\endmcitethebibliography\endthebibliography}{}

%  Fig1:    Charge distribution   ###################################### 

%  Fig2:  Quantum Espresso     ###################################### 

%     Fig3:  Plasmonics     ###################################### 

%      Fig4:  atomistic RDF NPs-HCQ ; NPs-CQ   ###################################### 

%      Fig5: RDF NPs-HCQ ; NPs-CQ   ######################################

%      Fig6:  RDF Ag-HCQ  Different sizes  ###################################### 

\end{document}